\documentclass[aps,twocolumn,prx,superscriptaddress,nofootinbib,amsmath,amssymb,floats,floatfix,english]{revtex4-1}
\usepackage{polski}
\usepackage[utf8]{inputenc}
\usepackage[T1]{fontenc}
\usepackage{amsmath}
\usepackage{amssymb}
\usepackage{ifthen}
\usepackage{graphicx}
\usepackage{times}
\usepackage{babel}
\usepackage{color}
\usepackage{bm}
\usepackage[normalem]{ulem}
\usepackage{float}
\usepackage{placeins}
\usepackage{chngcntr}
\usepackage{enumitem}
\usepackage{multirow}

\usepackage{comment}

\newcommand{\ifis}[2]{
\ifthenelse{\equal{#1}{}}{}{#2}
}

\def\be{\begin{equation}}
\def\ee{\end{equation}}
\def\bea{\begin{eqnarray}}
\def\eea{\end{eqnarray}}
\def\bsu{\begin{subequations}}
\def\esu{\end{subequations}}
\def\bi{\begin{itemize}}
\def\ei{\end{itemize}}

\newcommand{\op}[1]{\widehat{#1}}
\newcommand{\dagop}[1]{\widehat{#1}^{\dagger}}

\newcommand{\mc}[1]{{\mathcal{#1}}}
\newcommand{\wt}[1]{{\widetilde{#1}}}
\newcommand{\wb}[1]{{\overline{#1}}}

\newcommand{\nonu}{\nonumber}
\newcommand{\etal}{~\textsl{et al.}}

\newcommand{\bra}[1]{\langle#1\vert}
\newcommand{\ket}[1]{\vert#1\rangle}
\newcommand{\braket}[2]{\langle#1\vert#2\rangle}

\DeclareMathOperator{\TR}{Tr}

\newcommand{\Tr}[1]{\TR\left[{#1}\right]}
\newlength{\templength}

\newenvironment{eq}{\begin{equation}}{\end{equation}}

\newcommand{\eqn}[1]{(\ref{#1})}

\renewcommand{\eq}[2]{\begin{equation}\label{#1}#2\end{equation}}
\newcommand{\eqs}[2]{\begin{subequations}\label{#1}\begin{eqnarray}#2\end{eqnarray}\end{subequations}}

\newcommand{\eqa}[2]{\begin{eqnarray}\label{#1}#2\end{eqnarray}}

\usepackage{lmodern}

\setlist[itemize]{nosep}
\setlist[enumerate]{nosep,nolistsep}

\newcommand{\REM}[1]{\ifthenelse{0=1}{#1}{}}

\begin{document}

\title{Fully quantum scalable description of driven dissipative lattice models} 

\author{Piotr Deuar}
\email{deuar@ifpan.edu.pl}
\affiliation{Institute of Physics, Polish Academy of Sciences, Aleja Lotnik\'ow 32/46, 02-668 Warsaw, Poland}

\author{Alex Ferrier}
\affiliation{Department of Physics and Astronomy, University College London, Gower Street, London WC1E 6BT, United Kingdom}

\author{Michał Matuszewski}
\affiliation{Institute of Physics, Polish Academy of Sciences, Aleja Lotnik\'ow 32/46, 02-668 Warsaw, Poland}

\author{Giuliano Orso}
\affiliation{Universit\'e  de  Paris,  Laboratoire  Mat\'eriaux  et  Ph\'enom\`enes  Quantiques,  CNRS,  F-75013,  Paris,  France}

\author{Marzena H. Szyma\'nska}
\affiliation{Department of Physics and Astronomy, University College London, Gower Street, London WC1E 6BT, United Kingdom}

\date{\today}
\begin{abstract} 
Methods for modeling large driven dissipative quantum systems are becoming increasingly urgent due to recent experimental progress in a number of photonic platforms.  
We demonstrate the positive-P method to be ideal for this purpose across a wide range of parameters, focusing on the archetypal driven dissipative Bose-Hubbard model.  Notably, these parameters include intermediate regimes where interactions and dissipation are comparable, and especially cases with low occupations for which common semiclassical approximations can break down.  The presence of dissipation can alleviate instabilities in the method that are known to occur for closed systems, allowing the simulation of dynamics up to and including the steady state.  Throughout the parameter space of the model, we determine the magnitude of dissipation that is sufficient to make the method useful and stable, finding its region of applicability to be complementary to that of truncated Wigner. We then demonstrate its use in a number of examples with nontrivial quantum correlations, including a demonstration of solving the urgent open problem of large and highly non-uniform systems with even tens of thousands of sites. 
\end{abstract} 

\maketitle

\section{Introduction}
\label{INTRO}
Due to the rise in experimental progress with numerous photonic platforms, the dynamics and steady-state behavior of driven dissipative quantum systems \cite{Carusotto13} have received a great amount of both theoretical and experimental interest in recent times.  A variety of physical realizations, including cavity \cite{Raimond01r,Walther06r,Reiserer15r} and circuit QED systems \cite{Schmidt13,Houck12,Fink17,Fitzpatrick17,Kollar19},
arrays of coupled optical cavities \cite{Carusotto09,Umcal12} or of quantum dots \cite{Kasprzak2010}, hybrid systems \cite{Jin15}, 
polariton lattices \cite{Amo16r,Schneider16,Lai07,Kim11,Tanese2013,Tanese14,Zhang15,Baboux16,St-Jean17,Klembt17,Klembt2018,Whittaker18,Goblot19,Milicevic19,Su20,Dang20,Dusel20},  and certain implementations of ultracold atoms \cite{Brennecke2007}, can to varying degrees explore regimes in which both strong quantum correlations and dissipation to the environment are relevant effects.  

Unbiased quantum methods, including corner-space renormalization \cite{Finazzi15} and quantum trajectories \cite{Daley14,Biondi17b} can successfully treat small systems, but suffer from the usual runaway complexity problems once larger numbers of modes or sites are present. This issue is exacerbated even further for open systems since density matrices are needed, where the number of variables scales as $(e^M)^2$ with the configuration size $M$ rather than ``only'' $e^M$ for pure states. 
Matrix product states and related techniques \cite{Zwolak04,Biondi15} offer one way around this for closed systems, but their extension to include drive and dissipation is difficult \cite{Jin:PRL2013}. 

In contrast, techniques known as \mbox{\it phase-space methods}, in which quantum expectation values are calculated from averages over stochastic trajectories in phase-space, are readily scalable to quantum problems with large numbers of sites or modes, and are naturally adapted to open systems due to already being based on a density matrix formalism. Their performance does not depend much on dimensionality. Indeed, the use of the approximate truncated Wigner method has become 
common for studying semiclassical phenomena in ultra-cold atoms and microcavity polaritons \cite{Norrie05,Sinatra02,Martin10b,Steel98,Deuar09b,Hoffmann08,Carusotto05,Wouters09,Dagvadorj15,Dominici15,Donati16,Caputo17,Comaron18,Ballarini20,Zamora20}.  However, as lattice experiments increasingly aim to delve further into the quantum regime in these media, other techniques are needed to study quantum effects beyond the reach of the truncated Wigner approximation.

An alternative phase-space method, the positive-P approach \cite{Drummond80} allows for the full quantum mechanics of systems with up to two-body interactions to be simulated in an unbiased way without approximations. It has already found significant application in quantum optics \cite{Carter87}, and in ultracold atoms \cite{Lewis-Swan14}, where it has been successfully applied to cases with hundreds or even millions of sites \cite{Deuar07,Kheruntsyan12}. For closed systems, the trade-off has always been that while results for short evolution times are accessible, a nonlinear amplification of the trajectory spread eventually appears at sufficiently long times to obscure predictions below a rising noise floor \cite{Gilchrist97,Deuar06a}. 
However, it is already known that dissipation is beneficial to the stability of the method, and simulations can stabilize fully if it is sufficiently large \cite{Gilchrist97}.  

It is with this in mind, and with the increasing relevance of the physics of open quantum systems to a number of experimental platforms, that we propose positive-P as an ideal method for simulating such systems in intermediate regimes, relevant to current experiments, where driving, dissipation and quantum correlations are all relevant effects.  To demonstrate this, we focus on the archetypal driven dissipative Bose-Hubbard model, which is directly applicable to a number of the different experimental realizations \cite{Carusotto13,Naether15}.  We firstly endeavor to thoroughly characterize the regimes of applicability of positive-P in the parameter space of the driven dissipative Bose-Hubbard model, before also demonstrating a number of specific examples of nontrivial effects accessible to the method, some of which may be difficult to solve accurately by other means due to the very large or highly non-uniform systems considered.  The success of the positive-P method demonstrated here for the driven dissipative Bose-Hubbard model also implies that the stabilizing effect of dissipation on the method should likely allow it to be useful for simulating a number of related models of open quantum systems in future.  We also demonstrate that the regions of applicability of positive-P and truncated Wigner happen to be complementary to each other, with the truncated Wigner approximation being fairly accurate for large occupations (i.e. strong drive) and positive-P being stable for strong dissipation. Between them they provide a viable phase-space method for almost all regimes where external drive and/or dissipation are significant effects.  

\begin{figure*}[htb]
\begin{center}
\includegraphics[width=2.0\columnwidth]{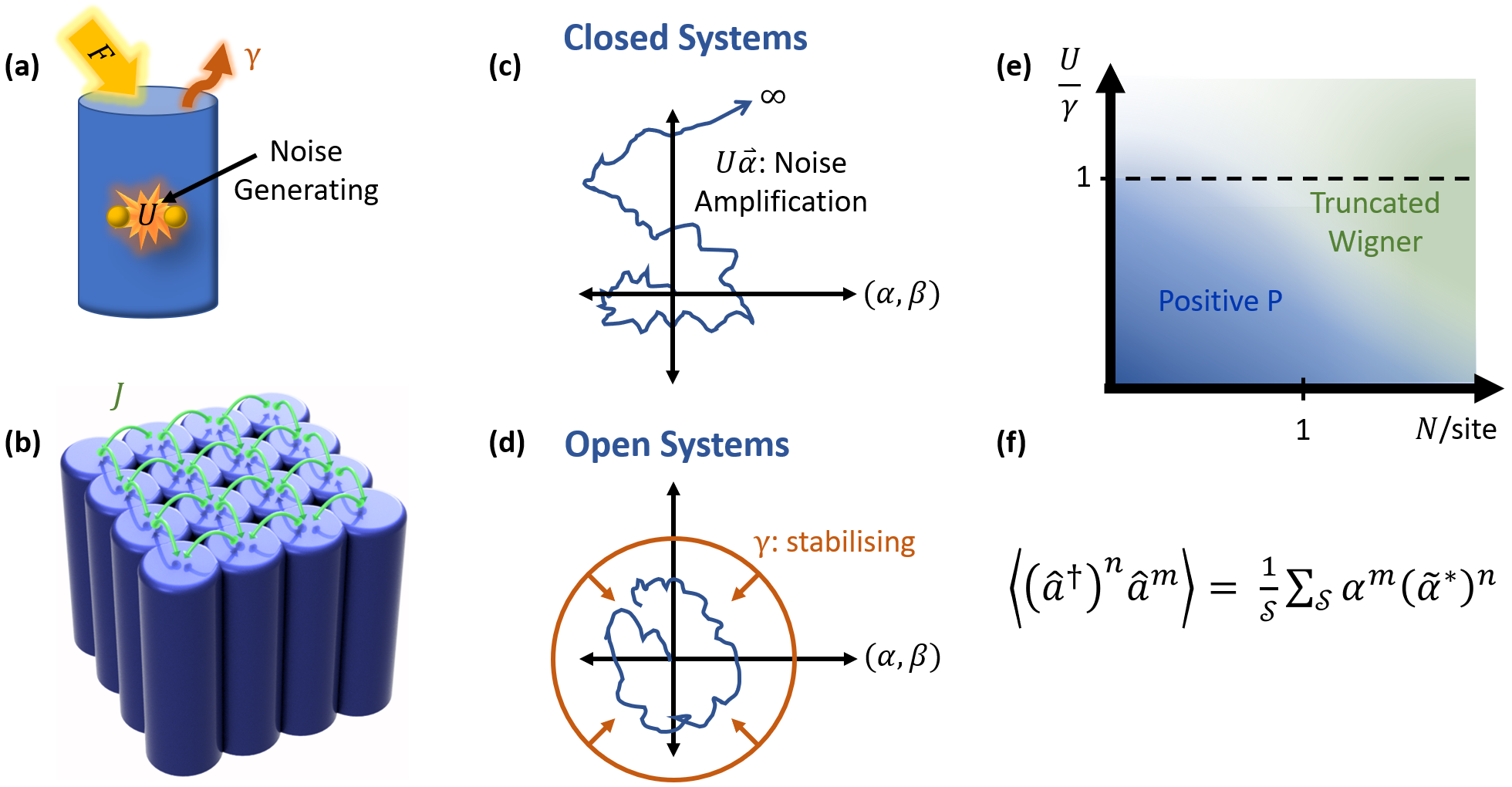}
\end{center}\vspace*{-0.5cm}
\caption{{\bf Illustration of the application of Positive-P to driven dissipative Bose-Hubbard models.}  (a) Sketch of the local processes involved in the model: external drive $F$, dissipation $\gamma$, and two-body interactions $U$. Only two body processes, such as the interactions $U$, generate the noise terms in the positive-P equations.  (b) Hopping $J$ couples connected sites in a lattice.  (c) In closed systems, noise amplification causes trajectories to escape to infinity in finite time.  (d) Sufficient dissipation can stabilize the trajectories, allowing the simulations to reach the steady state.  (e) Rough sketch of the regions of applicability of the positive-P and truncated Wigner methods in parameter space.  Positive-P works especially well for low occupations and/or strong dissipation, while the truncated Wigner approximation is accurate for large occupations (see Fig.~\ref{fig:TW}).  (f) In positive-P, normally ordered quantum observables are calculated by averaging the corresponding stochastic phase space variables over realizations.  This correspondence is exact in the limit of large numbers of realizations.  
\label{leadfig}}
\end{figure*}

The paper is organized as follows: 
In Sec.~\ref{MODEL} we describe the driven dissipative Bose-Hubbard model, and then in Sec.~\ref{PP} present its mapping to the positive-P representation \eqn{pp}.  
Sec.~\ref{1MODE} studies the single-site case and determines the level of damping \eqn{urule} needed for successful simulation, while benchmarking against known exact solutions.
We then investigate use cases in multimode models (Sec.~\ref{LATTICES}), including Lieb lattices with dark sites and large non-uniformly driven 2d square lattices, demonstrating scalability to huge systems (Fig.~\ref{fig:ifpan}).
Extension to nonzero temperature is given in Sec.~\ref{TEMP} before concluding in Sec.~\ref{CONC}.  An illustration indicating the key messages of this paper is presented in Fig. \ref{leadfig}.  

\section{Model}
\label{MODEL}
The Bose-Hubbard model is the standard go-to description for bosonic driven-dissipative lattice systems. In dimensionless units the Hamiltonian can be written in the tight-binding form:
\eq{H}{
\op{H}\ =\ \sum_j \op{H}_j\quad-\sum_{{\rm connections}\,i,j}\left[J_{ij}\dagop{a}_j\op{a}_i + J_{ij}^*\dagop{a}_i\op{a}_j\right].		
}
Here, the local part of the Hamiltonian at site $j$ is
\begin{equation}
\op{H}_j = -\Delta_j\dagop{a}_j\op{a}_j +\frac{U_j}{2}\,\dagop{a}_j\dagop{a}_j\op{a}_j\op{a}_j +F_j\dagop{a}_j + F_j^*\op{a}_j
\end{equation}
where $\op{a}_j$ is the bosonic annihilation operator at site $j$, $U_j\ge 0$ the local two-body interaction, $F_j$  the strength of coherent driving (can be complex), and $-\Delta_j$ is the local energy bias. 
For example, for polaritons in micropillars \cite{Tanese2013,Tanese14,Baboux16,St-Jean17,Klembt17,Whittaker18,Klembt2018,Goblot19,Milicevic19} with pumping frequency $\omega_p$ and natural mode frequency $\omega_j$, the  $\Delta_j = \omega_p-\omega_j$ plays the role of an effective chemical potential \cite{Biondi17b}.
Returning to \eqn{H}, $J_{ij}=J_{ji}^*$ is the tunneling amplitude for a transfer $i\to j$ between connected sites. 
For definiteness, in this notation, each connection occurs only once in the sum, so that e.g. a system consisting of just two connected sites has the tunneling terms $-J_{12}\dagop{a}_2\op{a}_1 - J_{12}^*\dagop{a}_1\op{a}_2$.
Complicated connections and lattices can also be trivially incorporated into the model via the general form in \eqn{H}.  While in this work we consider examples with nearest neighbour connections in one or two dimensions, there is no reason in principle that these methods should be any less effective for higher dimensions, all-to-all connections, or longer range tunneling that could be represented by arbitrary $J_{ij}$. 

The local single-particle dissipation rate is $\gamma_j$. The system is then described via the density matrix $\op{\rho}$ and evolves according to the master equation
\eq{drhodt}{
\frac{\partial\op{\rho}}{\partial t} = -i\left[\op{H},\op{\rho}\right] + \sum_j\frac{\gamma_j}{2}\left[2\op{a}_j\op{\rho}\,\dagop{a}_j -\dagop{a}_j\op{a}_j\op{\rho} - \op{\rho}\,\dagop{a}_j\op{a}_j\right].
}
This assumes dissipation into empty modes. The case of non-empty reservoir modes 
is described in Sec.~\ref{TEMP}.

For a single mode (site) with parameters $F$, $\Delta$, $U$, $\gamma$, the observables of most interest are the mode occupation
$N = \langle\dagop{a}\op{a}\rangle$, 
mean amplitude $\langle\op{a}\rangle$, and 
normalised two-body correlation
$g_2 = \langle\dagop{a}\dagop{a}\op{a}\op{a}\rangle/\langle\dagop{a}\op{a}\rangle^2$. 
Bunching is indicated by $g_2>1$ and antibunching by $g_2<1$. Strongly antibunched modes can in principle be good quantum sources of single photons. 
The steady state solution of the single mode has been calculated analytically by Drummond and Walls \cite{Drummond80b}.  
Several regimes can be identified based on which process is dominant on the observables $N$ and $g_2$:
\begin{enumerate}
\item A \textsl{strongly driven regime} when $|F|\gg U$ and $|F|\gg\gamma$ with coherent high occupation in the stationary state $N\approx(|F|/U)^{2/3}$, $g_2\sim 1$.
\item An \textsl{interaction dominated regime} when $U\gg|F|$ and $U\gg\gamma$ with low occupation $N\lesssim 1$ and strong antibunching $g_2\ll 1$.  
\item A \textsl{strongly damped regime} when $\gamma \gg U$ and $\gamma\gg|F|$. Here $N\approx (2|F|/\gamma)^2$, and $g_2\sim 1$.
\item Detuning can eventually dominate if it is strong enough and typically leads to lower occupations, according to $N\approx (|F|/|\Delta|)^2$ (though at small $\gamma$, much more complicated behavior appears \cite{Biondi17b}).
\end{enumerate}
Coupling different sites will inevitably mix the different regimes, leading to novel quantum phenomena \cite{LeBoite13,Biondi17b,Casteels17,Naether15}, including more exotic physics with hysteresis and large collective fluctuations  \cite{LeBoite14,Biondi17a,Biondi17b}. Needless to say, no exact solution of the steady state of the many-site problem is currently available, even in one dimension.  
Models with space-dependent parameters are certainly possible and often demonstrated experimentally (e.g. micropillars allow for the fabrication of systems with parameters that are very flexible from site to site \cite{Jamadi2020}), but have been much less studied and simulated. Time dependence is also possible -- most readily for $F(t)$. 

\section{positive-P representation}
\label{PP}

The application of the positive-P representation \cite{Drummond80} 
to the model \eqn{H}--\eqn{drhodt} 
generally follows the standard procedure applied to the related ultracold Bose gas systems without drive and dissipation \cite{Deuar06a,DeuarPhD}. 
One expresses the density matrix of an $M$ mode/site system as
\begin{eqnarray}
\op{\rho} = \int d^{2M}\pmb{\alpha}\,d^{2M}\wt{\pmb{\alpha}}\ P(\pmb{\alpha},\wt{\pmb{\alpha}}^*)\ \op{\Lambda}(\pmb{\alpha},\wt{\pmb{\alpha}}^*) \nonu \\
\op{\Lambda}=\bigotimes_j \op{\Lambda}_j(\alpha_j,\wt{\alpha}^*_j);\quad \op{\Lambda}_j = \frac{\ket{\alpha_j}_j\bra{\wt{\alpha}_j}_j}{\braket{\wt{\alpha}_j}{\alpha_j}} 
\end{eqnarray}
in terms of local coherent state kernels $\op{\Lambda}_j$ at each site $j$, with ${\rm Tr}[\op{\Lambda}_j]=1$. The $\ket{\alpha_j}_j$ and $\ket{\wt{\alpha}_j}_j$ are local coherent states 
$\ket{\alpha_j}_j = \exp\left[\alpha_j\dagop{a}_j\right]\,\ket{\rm vac}.$
The bold notation $\pmb{\alpha}$ indicates a vector of all $\alpha_j$ values. 
As a result of the properties of $\op{\Lambda}$, the distribution $P$ can be made positive real for any density matrix, hence it is a true probability distribution of the configurations $\vec{v}=\{\pmb{\alpha},\wt{\pmb{\alpha}}^*\}$ \cite{Drummond80}.
For this to be possible, however, the $\wt{\alpha}_j$ ``bra'' duals to the ``ket'' amplitudes $\alpha_j$ must be independent, leading to an off-diagonal kernel operator $\op{\Lambda}_j$. 
There is a full equivalence between the density matrix $\op{\rho}$ and the distribution $P(\vec{v})$. Moreover, a set of $\mc{S}$ samples of the configuration $\vec{v}$, distributed according to $P$, is also equivalent to the full density matrix in the limit $\mc{S}\to\infty$. Therefore, a set of such samples can in principle be used to approximate full quantum mechanics with increasing and unbiased precision as $\mc{S}$ grows. 

We can then use the properties of the projector $\op{\Lambda}$ to convert the master equation \eqn{drhodt} into a Fokker-Planck equation for the evolution of the distribution $P$ (see Appendix \ref{DevPP} for details), which in turn leads to stochastic differential equations for trajectories of the phase space variables $\vec{v}$. 
The resulting (It\^o) stochastic equations for the samples of $\vec{v}$ are
\eqs{pp}{
\frac{\partial\alpha_j}{\partial t} &= i\Delta_j\alpha_j -iU_j\alpha_j^2\wt{\alpha}^*_j -iF_j -\frac{\gamma_j}{2}\alpha_j \notag \\
& + \sqrt{-iU_j}\,\alpha_j\,\xi_j(t) + \sum_{k} iJ_{kj}\alpha_{k}, \label{ppa}\\
\frac{\partial\wt{\alpha}_j}{\partial t} &= i\Delta_j\wt{\alpha}_j - iU_j\wt{\alpha}_j^2\alpha_j^* -iF_j -\frac{\gamma_j}{2}\wt{\alpha}_j \notag \\
& + \sqrt{-iU_j}\,\wt{\alpha}_j\,\wt{\xi}_j(t) + \sum_{k} iJ_{kj}\wt{\alpha}_{k} \label{ppb}
}
where the final sum is over all sites $k$ connected to $j$.
The real random variables $\xi_j(t)$ and $\wt{\xi}_j(t)$ are independent white noises of mean zero obeying \mbox{$\langle\xi_j(t)\xi_{k}(t')\rangle_s = \delta(t-t')\delta_{jk}$,} \mbox{$\langle\wt{\xi}_j(t)\wt{\xi}_{k}(t')\rangle_s = \delta(t-t')\delta_{jk}$,} and \mbox{$\langle\xi_j(t)\wt{\xi}_{k}(t')\rangle_s = 0$,} where the notation $\langle\cdot\rangle_s$ denotes stochastic averaging over the available samples in the limit $\mc{S}\to\infty$.  Moreover, $\xi_jdt$ and $\tilde \xi_jdt$  are standard Wiener increments, which are implemented by Gaussian random variables of variance $1/\Delta t$ at each time step of length $\Delta t$.

The equations \eqn{pp} are the ones to be solved numerically and our subsequent analysis in this paper is based upon them. They contain the full quantum mechanics of the system, provided that the noise amplification catastrophe alluded to above does not occur (The useful simulation time $t_{\rm sim}$ beforehand is estimated in Appendix \ref{SIMT}).

\section{Single-mode performance}
\label{1MODE}

Let us start with the baseline single mode case, because it is very revealing regarding the capabilities of the method, and allows us to easily compare to the exact solution, as was given by Drummond and Walls \cite{Drummond80b}. It also turns out to be an excellent guide for assessing which many-site systems can be simulated, and lets us understand more involved multi-mode systems that will follow. 
We omit the site indices $j$ in this Section. 
The observables of most interest have the following stochastic estimators in the positive-P calculations:
\begin{equation}\label{obs}
N = \langle\dagop{a}\op{a}\rangle = {\rm Re}\langle(\alpha\wt{\alpha}^*)\rangle_s, \quad \langle\op{a}\rangle = \langle\alpha\rangle_s = \langle\wt{\alpha}\rangle_s, 
\end{equation}
\begin{equation}\label{obs2}
g_2 = \frac{\langle\dagop{a}\dagop{a}\op{a}\op{a}\rangle}{\langle\dagop{a}\op{a}\rangle^2} = \frac{{\rm Re}\langle(\alpha\wt{\alpha}^*)^2\rangle_s}{N^2}.
\end{equation}
For all results we present in this work, we begin simulations in vacuum ($\alpha=\wt{\alpha}=0$) and evolve until the steady state is reached (or until excessive noise amplification makes further simulation pointless). 
Appendix \ref{SIMT} gives a perspective on other initial states.

\subsection{Regimes of usefulness} 
\label{USEFUL}

\begin{figure}[htb]
\begin{center}
\includegraphics[width=\columnwidth]{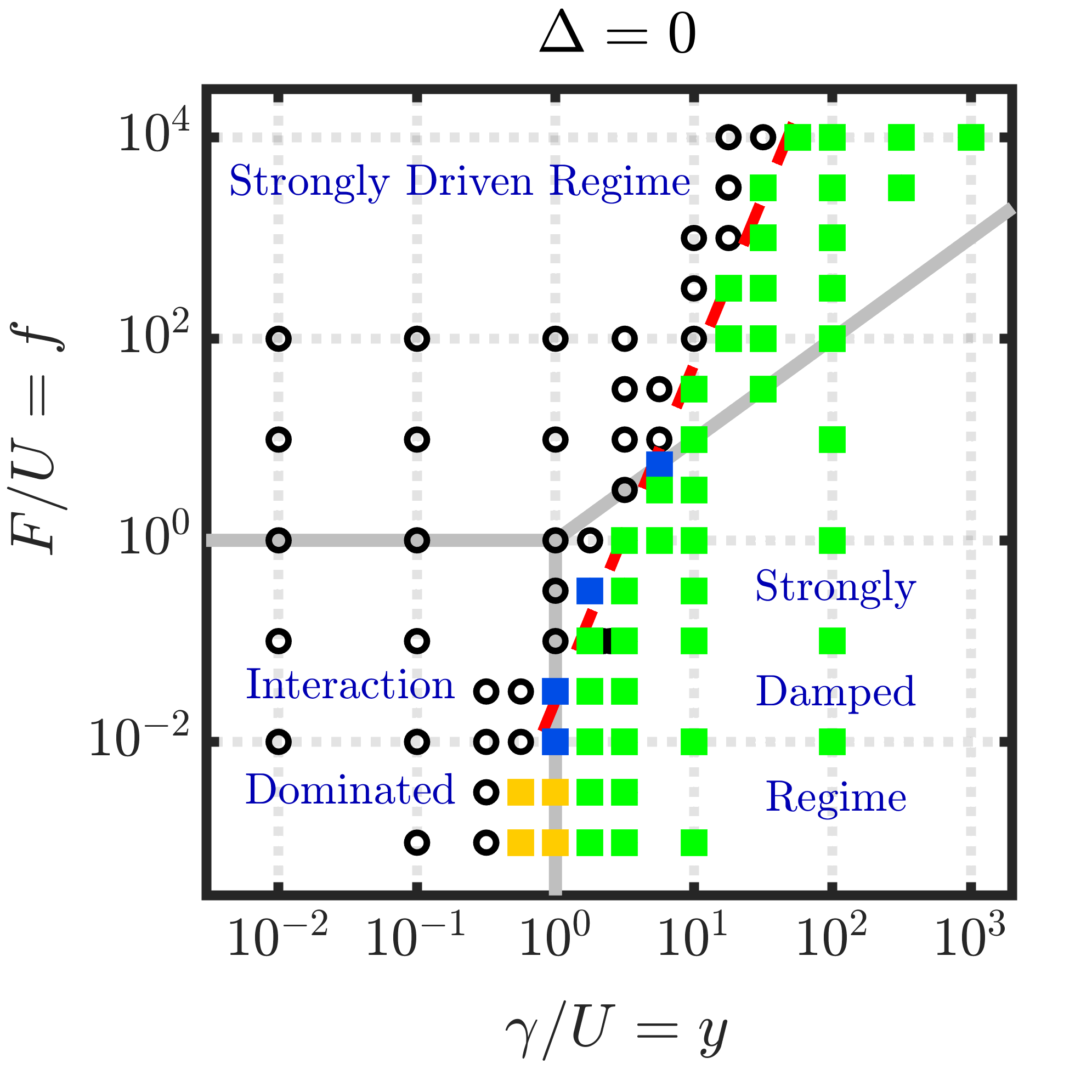}
\end{center}\vspace*{-0.5cm}
\caption{
{\bf Regimes of usefulness of positive-P numerical calculations}. Symbols show the performance on the 1-mode model when $\Delta=0$. 
Green square: numerics reaches the stationary state and remains stable;
Yellow square: remains stable, but poor signal-to-noise ratio makes accurate determinations intractable (especially for $g_2$); 
Blue square: numerics reaches the stationary state but does not remain stable later;
Open circle: numerics becomes unstable before reaching the stationary state. 
The broad grey lines indicate crossovers between physical regimes listed in Sec.~\ref{MODEL}; the red dashed line shows the empirical estimate of the usability region \eqn{usability}.
\label{fig:phasediag}}
\end{figure}

\begin{figure*}[htb]
\begin{center}
\includegraphics[width=2.05\columnwidth]{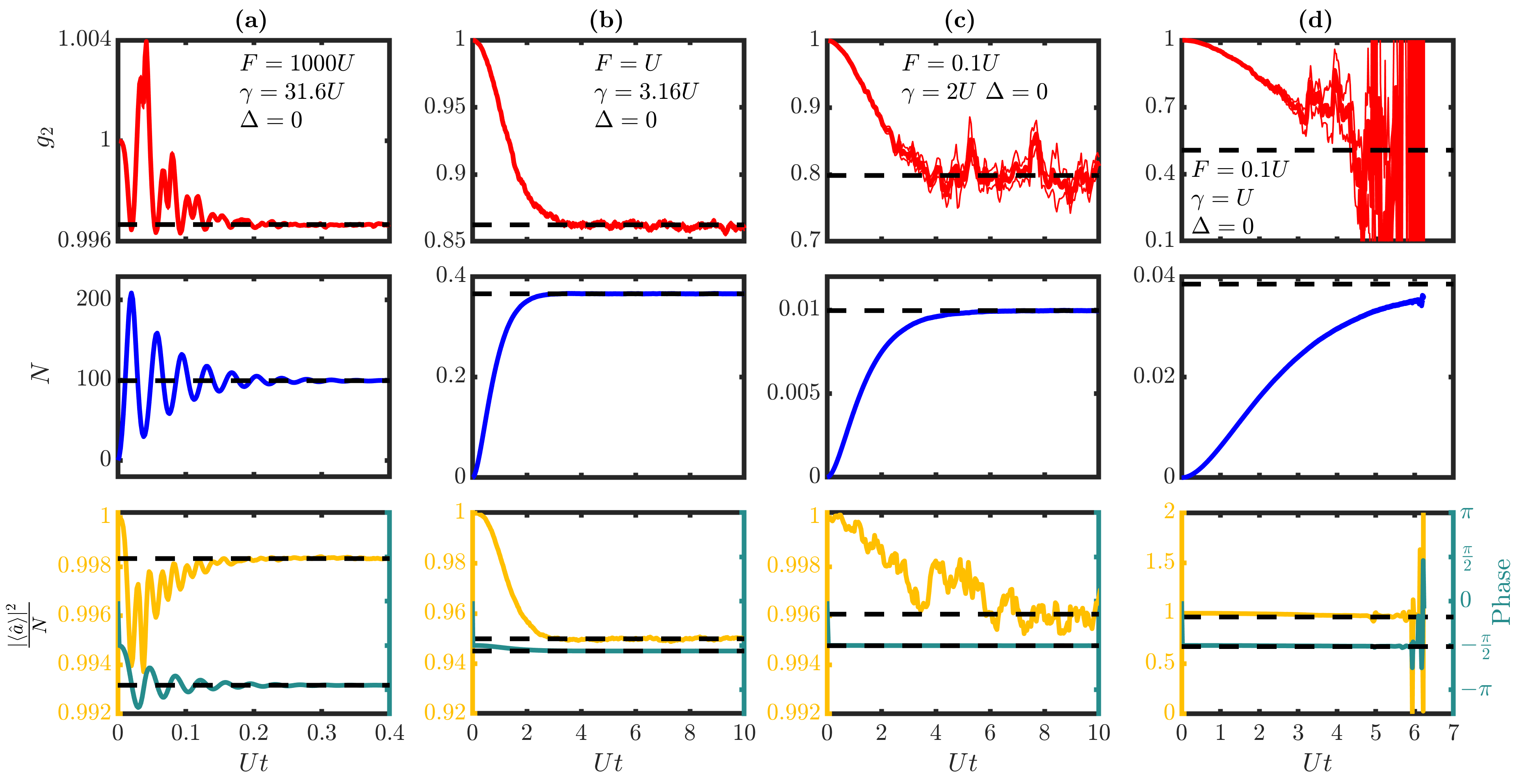}
\end{center}\vspace*{-0.5cm}
\caption{ {\bf Examples of numerical integration with Positive-P.}
 $g_2$ (red, top row), $N$ (blue, middle row), phase (green) and square amplitude normalized by $N$ (yellow) of $\langle\hat{a}\rangle$ (bottom row) for different regimes in the one site system. Column (a)  $F = 1000U$, $\gamma=31.6\,U$, $\Delta=0$ in the strongly driven regime, column (b) $F=U$ and $\gamma=3.16\,U$ in a crossover regime, column (c) $F = 0.1\,U$, $\gamma=2\,U$, $\Delta=0$ in the weakly pumped regime with strong antibunching, column (d) $F = 0.1\,U$, $\gamma=U$, $\Delta=0$ interaction dominated regime with insufficient damping to reach the steady state before the noise instability occurs. This is an example of Positive-P failing. 
Solid colored lines: positive-P simulation, $10^6$ samples. 
Dashed black line: exact value \cite{Drummond80b}.
\label{fig:1m}}
\end{figure*}

A basic starting question is whether the stationary state can be reached.
For many-site systems, a rough minimum requirement is that single site simulations can do so -- under all the local conditions found in the large system.
Hence, the fundamental importance of determining the conditions under which a single site system can reach the stationary state.
We have carried out positive-P calculations across the whole spectrum of parameters for the single site system, and assessed them according to whether a stationary state with useful signal-to-noise ratio is reached. That is, whether for practical numbers of realisations, the values of the observables we consider in the steady state are not masked due to the self-amplification of the noise.  $U$ was chosen as an arbitrary energy scale.  Fig.~\ref{fig:phasediag} presents the results of this benchmarking, over many orders of magnitude of the parameters $F$, $U$, and $\gamma$, when $\Delta=0$.  This is one of the main results of the paper.

The stable region in which numerical integrations reach the steady state and remain well behaved is shown in green, and is attained for all parameters $F,U,\Delta$ when the damping $\gamma$ becomes sufficiently large.  
Examples of such calculations are shown in Fig.~\ref{fig:1m}(a, b, c).
Dynamics that do not reach the steady state before the noise instability occurs, such as Fig.~\ref{fig:1m}(d), are shown as a small open circle.  
The blue squares are on the edge of stability, such that a stationary state is reached, but noise instability similar to that shown in Fig.~\ref{fig:1m}(d) sets in some time after.  
The yellow square cases are stable, but mode occupation is too low compared to the vacuum noise, and useful information cannot be extracted.
We find that for nonzero $\Delta$, the regime of stability is qualitatively almost identical to that in Fig.~\ref{fig:phasediag}, particularly on a log-log scale (see Appendix~\ref{ADELTA} for details). 
Dependence on $\Delta$ is investigated further in Sec.~\ref{DETUNING}.

An empirical rule that largely captures the regime of usability, based on 
the data shown in Fig.~\ref{fig:phasediag}, is:
\label{urule}
\eq{usability}{
\gamma \gtrsim 3 U\ \left(\frac{F}{U}\right)^{0.30}.
}
The uncertainty is about $\pm0.01$ on the exponent, and 10\% on the prefactor.
For very low driving, a more appropriate rule is
\eq{usability1}{
\gamma \gtrsim U {\rm\qquad when\  }F\lesssim0.01U.
}
In the usable regime, numerical effort scales linearly with the number of sites, and quadratically with the precision (according to the central limit theorem, since all samples have independent noise input).
Much lower damping may become accessible through the use of stochastic gauges, particularly in the high occupation regime where they were shown to be effective for this Hamiltonian \cite{Deuar06b}.

\subsection{Typical behavior}
\label{BEHAVIOR}

Here, we now look in more detail at specific examples presented in Fig.~\ref{fig:1m}.  The case in Fig.~\ref{fig:1m}(a) with $F = 1000U$, $\gamma=31.6U$ and $\Delta=0$ is representative of the strongly driven regime, with a few oscillations before settling down to a steady state with high occupation and almost perfect coherence ($g_2\approx 1$).  
The crossover regime that mixes all three regimes mentioned in Sec.~\ref{MODEL}, and is often studied \cite{Casteels17,Casteels16,Biondi17a,Biondi17b,Finazzi15,LeBoite13}, is 
shown in Fig.~\ref{fig:1m}(b). There $F=U$, $\gamma=3.16U$, $\Delta=0$, and occupation is $\mc{O}(1)$. 
This case is notable in that we can obtain large antibunching, indicating strong quantum effects, while remaining stable and despite rather strong dissipation. 
Getting into lower occupations and stronger antibunching, Fig.~\ref{fig:1m}(c) shows the case of $F=0.1U$, $\gamma=2U$, $\Delta=0$. Notice that the statistical error in $g_2$ is becoming more pronounced, despite averaging over $10^6$ trajectories. This is still a well behaved simulation, however, without significant noise amplification. The fairly low signal to noise ratio is a consequence of low occupation.
When damping is insufficient to stabilize the long time behavior, a case like Fig.~\ref{fig:1m}(d) occurs, here with $F=0.1U$, $\gamma=U$, $\Delta=0$. 
The exact stationary value is approached, but the evolution does not convincingly stabilize before noise amplification appears (first spiking near $Ut\approx5$) and leads to an instability ($Ut\approx6.2$).  
As is common for higher order moments, the $g_2$ estimation becomes too noisy to be useful some time before.

\subsection{Nonzero detuning}
\label{DETUNING}

\begin{figure}[htb]
\begin{center}
\includegraphics[width=\columnwidth]{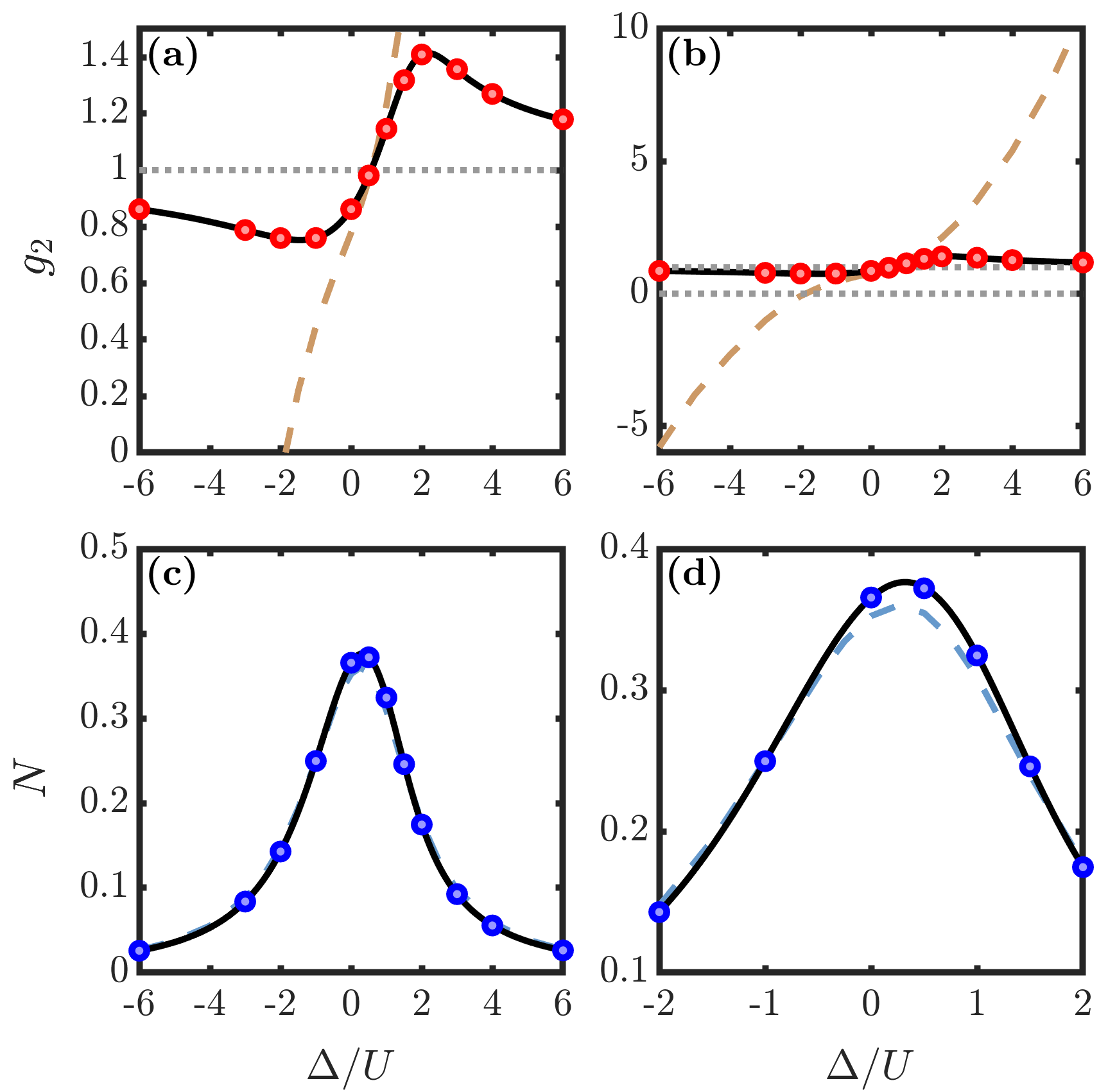}
\end{center}\vspace*{-0.5cm}
\caption{ {\bf Detuning dependence and comparison between Positive-P and Truncated Wigner.}
Variation of $g_2$ (a, b) and $N$ (c, d) with detuning $\Delta$ at $F = U$, $\gamma = 3.16U$; 
comparison of positive-P (circles) and truncated Wigner simulations (dashed) with exact results \cite{Drummond80b} (solid line). Panels (b, d) are the same data as (a, c) respectively, but with scale adjusted to display  deviation of truncated Wigner from exact and positive-P results.
\label{fig:deltaw}}
\end{figure}

While detuning $\Delta$ does not appreciably influence the regime of stability, 
the physics is significantly affected. 
Fig.~\ref{fig:deltaw}(a,c) shows the variation of occupation and bunching when $F=U$, $\gamma=3.16U$, close to the mixing region of all three regimes. This is quite a strongly damped case compared to many theoretical studies, but still shows strong bunching and antibunching. The form of this variation of $g_2$ with $\Delta$ is qualitatively consistent with the phase diagram calculated in the weakly damped $\gamma=0.05U$ regime \cite{Biondi17b}, just with a reduced degree of bunching/antibunching owing to the stronger dissipation relative to $U$.
Comparison with exact results shows that all detunings can be reliably and stably simulated, even close to the $\Delta=0$ limit of stability \eqn{usability}. 

The fact that those simulations remain stable for $\Delta\neq0$ can be attributed to the decreasing occupation.  
To understand this, note first that for the undamped system, single particle energy shifts of the kind represented by $\Delta$ were shown not to affect the stable simulation time given by \eqn{tsim}, for set values of $U$ and mean particle number $N$ \cite{Deuar06a}. In the damped system, a similar indirect-only dependence is expected, but $N$ does depend on $\Delta$. Since $|\Delta|>0$ generally reduces the particle number (Fig.~\ref{fig:deltaw}), the estimate \eqn{tsim} indicates increased $t_{\rm sim}$, so one expects increased stability and smaller $\gamma$ values than in Fig.~\ref{fig:phasediag} to become accessible.

\begin{figure}[htb]
\begin{center}
\includegraphics[width=\columnwidth]{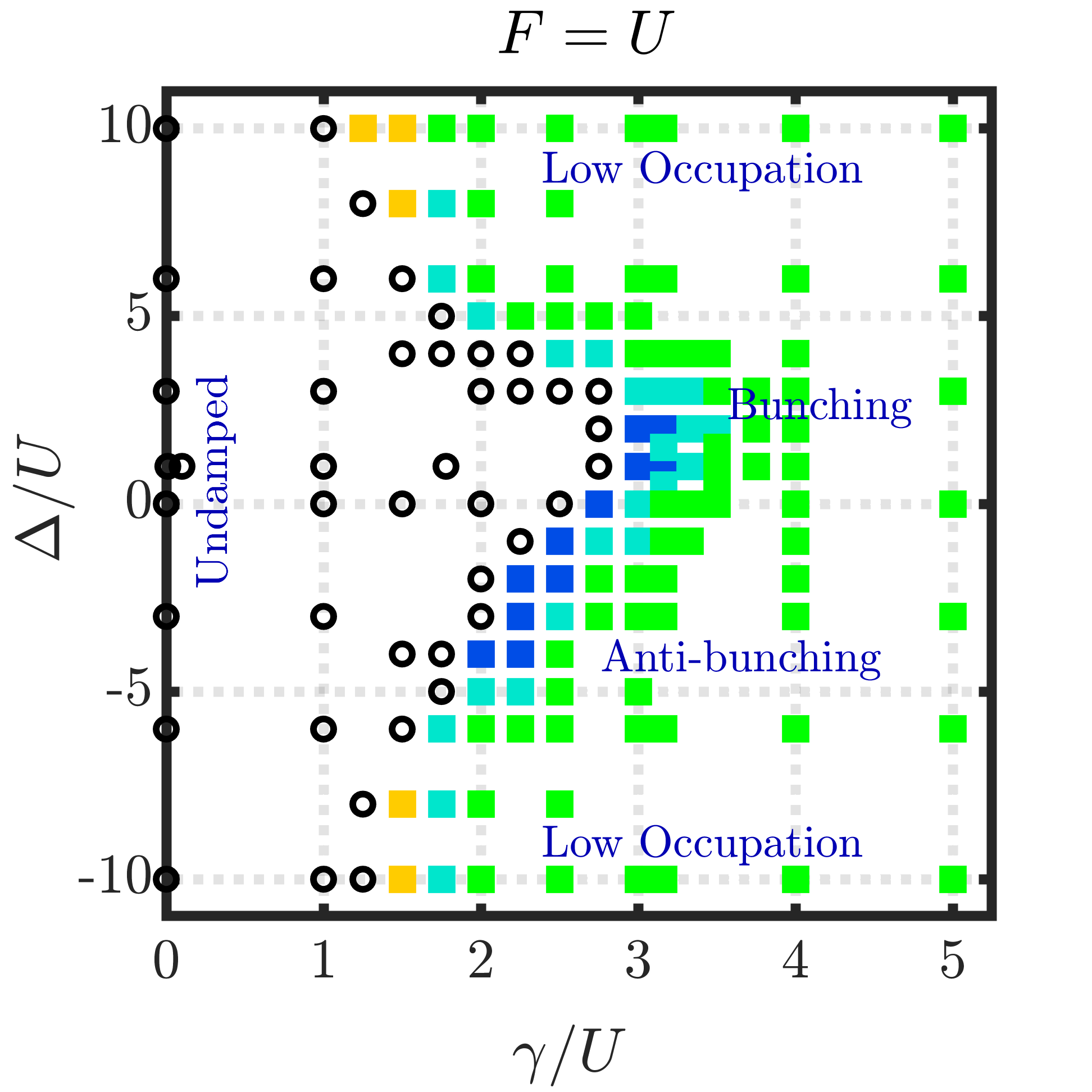}
\end{center}\vspace*{-0.5cm}
\caption{ {\bf Regimes of usefulness of positive-P as a function of detuning $\Delta$}. 
Notation the same as in Fig.~\ref{fig:phasediag}, with the addition of cyan cases when the instability after the steady state is reached was seen only for very large ensembles ($\mc{S}=10^6$) but not seen in $\mc{S}=10^5$ ensembles. 
\label{fig:phasediagdelta}}
\end{figure}

This is borne out in Fig.~\ref{fig:phasediagdelta}, which shows a study of this stability dependence at $F=U$, the typical case of interest.
Near the edge of the stable region, 
only very rare trajectories exhibit instability, such that small ensembles are usually still well behaved (cyan color in Fig.~\ref{fig:phasediagdelta}). 
Such a trade-off between better precision in larger ensembles, but encountering instability if one generates too many trajectories,
is typical for the positive-P method in borderline stable/unstable regimes. 

The unstable region is more asymmetric around  $\Delta=0$ than the density in Fig.~\ref{fig:deltaw}. Bunching correlates with increased number fluctuations, such that maximum excursions of occupation are larger for $\Delta>0$ than for $\Delta<0$, making instability persist at $\Delta>0$ for larger damping.  

\subsection{Comparison to truncated Wigner}
\label{WIGCOMP}

Fig.~\ref{fig:deltaw} compares the positive-P and exact results to those of a leading competitor for scalable quantum simulations -- the truncated Wigner approach (TW). 
This method's equations are described in Appendix~\ref{WIG}.  Unlike the positive-P method, truncated Wigner involves an approximation, namely that the exact evolution equation for the Wigner distribution (the equivalent of \eqn{ppFPE} for that representation) contains third order derivative terms, which must be neglected in order to obtain a stochastic differential equation from the resulting Fokker-Planck equation.  Physically, this means that some quantum correlations are not included in the description, which becomes an issue when looking at problems with a higher degree of entanglement and low mode occupations.  In other words, the more semiclassical the problem is, the better it is described with the truncated Wigner approach, which fails for very {\it quantum} cases.  
It is therefore useful to show that the positive-P may be applicable in situations where the truncated Wigner approximation fails to give accurate results, as well as compare their properties under conditions where either method would be viable. 

One can see that while truncated Wigner gives a qualitatively good description of the occupation (though with some deviations), two-body correlations $g_2$ are on the whole completely inaccurate. 
Unphysical predictions of $g_2<0$ with the truncated Wigner method are also seen. 
These 
are typical known problems with the truncated Wigner approach when occupations are low. The method gives much more accurate results for high occupations, such as in the strongly driven regime. Table~\ref{tab:snr} gives examples of this behavior for some other values of the parameters. 

For both methods, the statistical uncertainty on the steady state result is obtained by partitioning the \mbox{$\mc{S}\approx 10^6$} trajectories  
into (roughly) $s\sim 100$ subensembles, each containing $\mc{S}/s$ trajectories. For each subensemble $i$, we extract 
the steady state value $O_i$ of a given observable $\hat O$. We then consider these $s$ values as independent measurements, so that 
our best estimate is given by $O \pm \delta_{\rm{stat}} O$, where $ O=(\sum_i O_i)/s$ is the mean
and $\delta_{\rm{stat}} O=\sqrt{{\rm var}[O_i]/(s-1)} $ is the associated statistical error.    

The positive-P and truncated Wigner also differ with regard to the signal-to-noise ratio (SNR). At low occupations, the SNR is far superior in positive-P, while at high occupations it is comparable. This can be seen very clearly in Table.~\ref{tab:snr}.  
On the other hand, despite systematic errors and SNR issues, the truncated Wigner never suffers from the noise catastrophe of Fig.~\ref{fig:1m}(d).  

\begin{table}[htb]
\begin{center}
\begin{tabular}{|l|l|l|l|l|}
\hline
$F/U$							\quad 	&	1		\quad &	1		\quad &	0.01		\quad &	1000	\quad 	\\
$\gamma/U$						\quad 	&	3.16	\quad 	&	3.16	\quad 	&	2.0	\quad 	&	31.6	\quad 	\\
$\Delta/U$						\quad 	&	0		\quad &	-10	\quad 	&	0	\quad 	&	0	\quad 	\\
\hline
\quad$N$:						\quad 	&			&			&			&			\\
exact \cite{Drummond80b}	\quad 	&	.36589	\quad 	&	.0097392     \quad &	.000099996	\quad &	99.33055	\quad \\
Positive-P						\quad 	&	.3658(1) \quad 	&	.009741(2)	\quad &	.00009995(5)	\quad &	99.3305(8)	\quad \\
tr. Wigner						\quad 	&	.3525(2)	 \quad    & 	.01015(15)	\quad &	.00010(8)	\quad &	99.3309(5)	\quad \\	
\hline
\quad$g_2$:						\quad 	&			&			&			&			\\
exact \cite{Drummond80b}	\quad 	&	.86243	\quad &	\quad.90930	\quad &	\quad.799984		\quad &	.9966697	\quad \\
Positive-P						\quad 	&	.8628(6)	\quad &	\quad.9093(5) \quad &	\quad.801(5) \quad 	&	.996675(9)	\quad \\
tr. Wigner						\quad 	&	.779(2)	\quad &	-12(2)		\quad &	$\pm10^4$	\quad &	.996657(3)	\quad \\	
\hline
\end{tabular}
\end{center}\vspace*{-0.5cm}
\caption{
Comparison of stationary values from positive-P and truncated Wigner, with statistical uncertainty. All simulations used $10^6$ trajectories.
\label{tab:snr}}
\end{table}

\begin{figure}[htb]
\begin{center}
\includegraphics[width=\columnwidth]{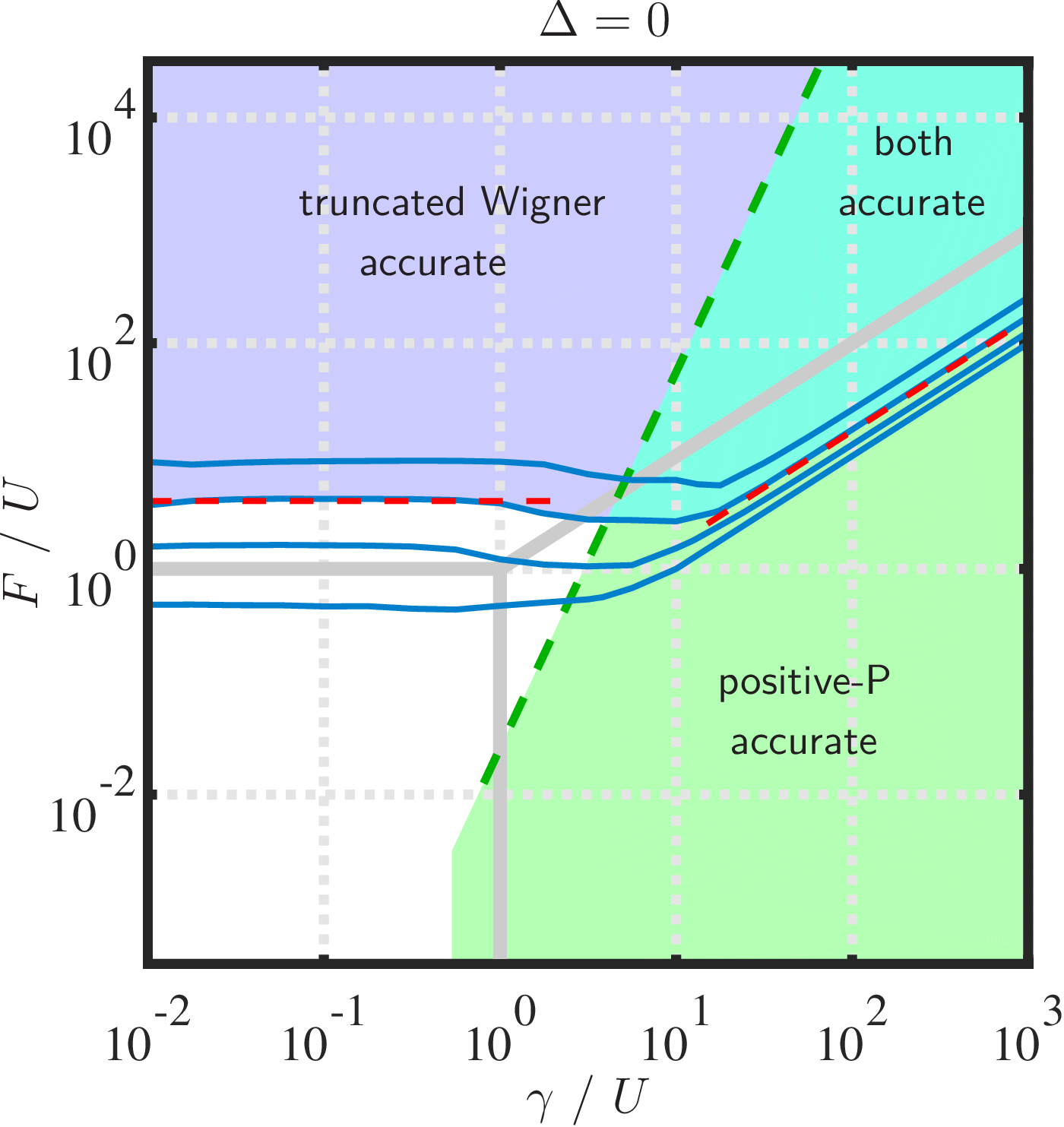}
\end{center}\vspace*{-0.5cm}
\caption{ 
{\bf Applicability of truncated Wigner and positive-P.} 
TW is assessed based on a figure of merit $\Delta_{TW}$ which characterizes the relative systematic and statistical errors of observables in the stationary state, as described in the text and \eqn{DeltaTW}. 
Blue lines show contours of $\Delta_{TW}=0.01, 0.03, 0.1, 0.3$ (top to bottom), dashed red lines the limits \eqn{urulewig}. For positive-P, the limits \eqn{urule} are used, \eqn{usability} shown as green dashes. Other notation follows Fig.~\ref{fig:phasediag}.
\label{fig:TW}}
\end{figure}

A systematic comparison of the applicability of the two methods is made in Fig.~\ref{fig:TW}, using the $\Delta=0$ case.  We assess the accuracy of the truncated Wigner by calculating both the systematic and statistical \emph{relative} errors, defined as $|O-O_{\rm ex}|/|O_{\rm ex}|$ and  $ \delta_{\rm{stat}} O/|O|$, respectively, for each of the four observables shown in Fig.~\ref{fig:1m}: $N$, $g_2$, $|\langle \op{a}\rangle|^2/N$ and phase $\arg\langle\op{a}\rangle$. Here  $O_{\rm ex}$ is the corresponding value of the observable obtained from the exact solution of Ref. \cite{Drummond80b}.  We then define $\Delta_{TW}$ as the maximum relative error out of the entire set (see Appendix~\ref{WIG} for details).  
The blue contours in Fig.~\ref{fig:TW} correspond to values of $\Delta_{TW}= 0.01, 0.03, 0.1, 0.3$ from top to bottom. We take $\Delta_{TW}=0.03$ as the nominal limit of sensible applicability of the truncated Wigner.  This curve determines the upper filled region in Fig.~\ref{fig:TW}, corresponding to the model
parameters that can be accurately simulated by the method.  

Explicit conditions for the TW accuracy region can be obtained by fitting the $\Delta_{TW}=0.03$ curve, in the 
asymptotic regimes of weak and of strong dissipation, to a straight line.  The obtained results are shown as red dashed lines in Fig.~\ref{fig:TW}, from which we find
\eqs{urulewig}{
  F \gtrsim 4U&\qquad&\text{ when $\gamma\lesssim 2U $}\label{urulewig1}\\
  F \gtrsim \gamma/6&\qquad&\text{ when $\gamma\gtrsim 20U $.}
}
The lower filled region in Fig.~\ref{fig:TW} refers instead to the regime of sensible applicability of the positive-P, where the empirical limits \eqn{urule} are used.  

The bottom line of this comparison is that the regimes of applicability of the positive-P and truncated Wigner methods are mostly complementary. 
TW is sufficient for small damping, high driving (alternatively -- large $N$), where the system behaves largely semi-classically and quantum correlations are small, while positive-P should be a method of choice for low driving, appreciable damping (low and moderate $N$) where quantum correlations are significant. 
Both are good in the high damping, high occupation regime.  
Together, these two phase space approaches cover the vast majority of the parameter space.  What is left is the low occupation (the limit in \eqn{urulewig1} with $N\approx(F/U)^{2/3}$ gives $N\approx 2.5$), low damping regime, which fortunately suits tensor network methods best.

\section{Lattices and multi-site systems}
\label{LATTICES}

\subsection{Unconventional photon blockade}
\label{BLOCKADE}

\begin{figure}[htb]
\begin{center}
\includegraphics[width=\columnwidth]{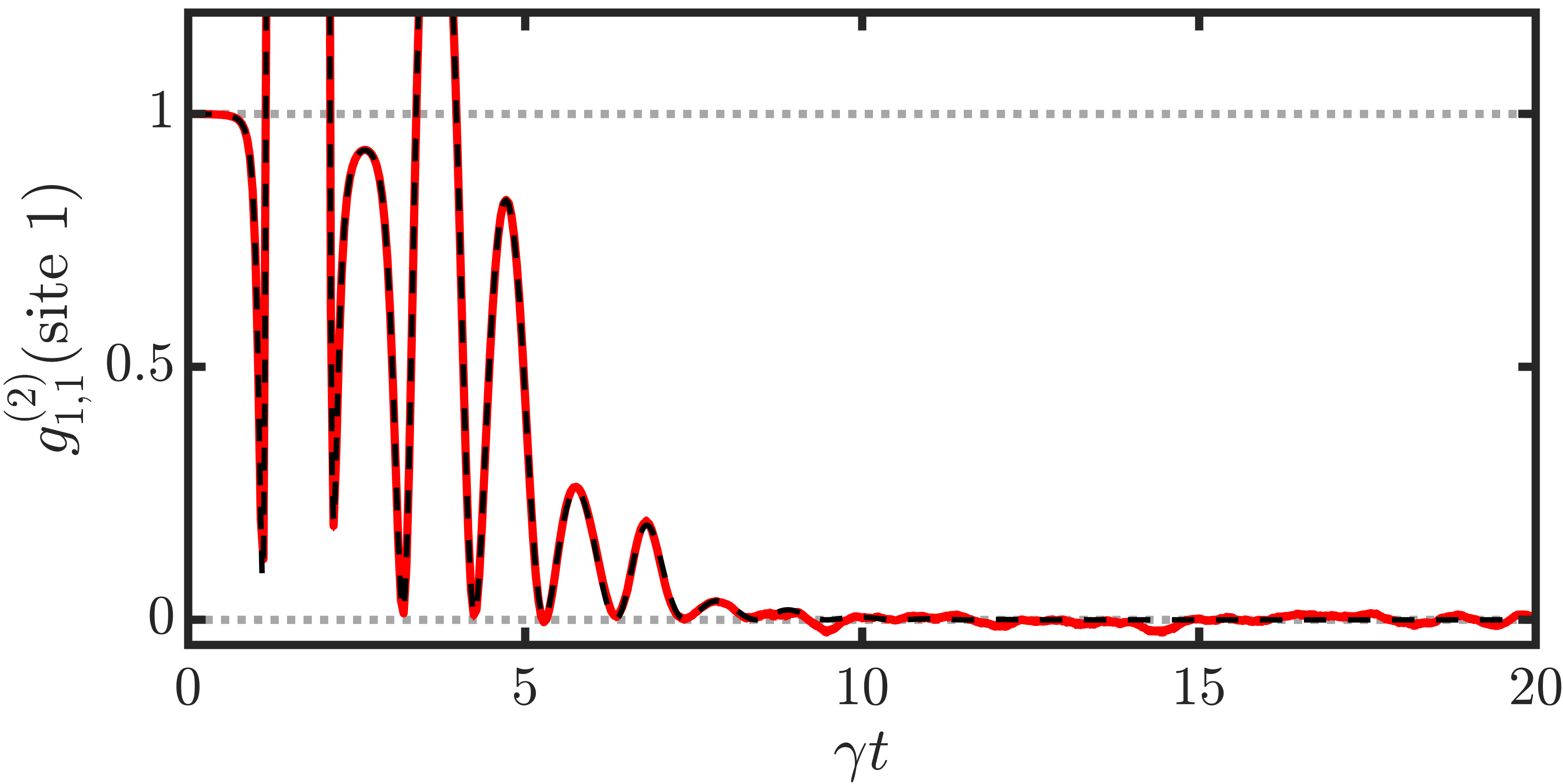}
\end{center}\vspace*{-0.5cm}
\caption{ {\bf Photon Blockade.}
A time trace of the density self-correlation, indicating strong unconventional photon blockade in a two-site system.  Positive-P (red) is compared with numerical exact solution of the master equation (dashed black). Parameters from \cite{Bamba11}, here $U=0.0856$, $J_{12}=3$,  $\gamma=1$, $\Delta=-0.275$ on two sites with driving $F=0.01$ on site 1 and $F=0$ on site 2. $\mc{S}=10^6$. We show the self-correlation of the driven site 1, $g^{(2)}_{1,1}=\langle\dagop{a}_1\dagop{a}_1\op{a}_1\op{a}_1\rangle/\langle\dagop{a}_1\op{a}_1\rangle^2$. 
\label{fig:blockade}}
\end{figure}

\begin{figure}[htb]
\begin{center}
\includegraphics[width=\columnwidth]{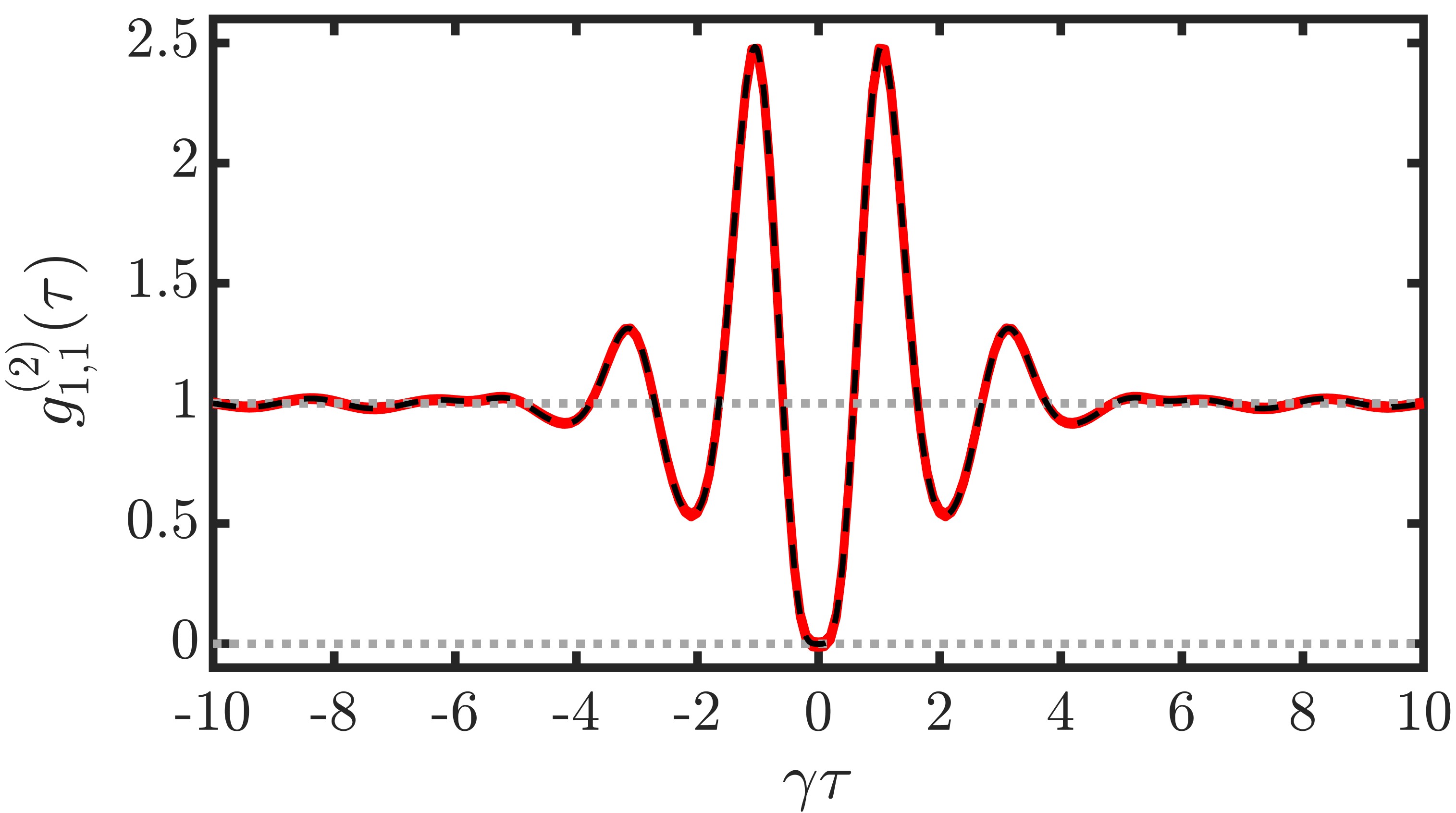}
\end{center}\vspace*{-0.5cm}
\caption{ {\bf Two time density correlations.}  Two time correlations $g^{(2)}_{1,1}\!\left(\tau\right)$ for the driven site in the steady state ( the same example as in Fig. \ref{fig:blockade}).   Positive-P (red) is compared with numerical exact solution of the master equation (dashed black). 
\label{fig:g2tau}}
\end{figure}
 
For the first many-mode example, we consider a situation where nontrivial behavior can occur in a system of only two sites.  Strong two-particle interference effects leading to $g_2\to0$ pose no problem to simulate. A calculation of the so-called unconventional photon blockade \cite{Liew10} (using parameters from \cite{Bamba11}) proceeds easily, as shown in Fig.~\ref{fig:blockade}. The steady state value obtained with $10^6$ 
realizations is $-0.001\pm0.004$.
This system consists of two sites ``1'' and ``2'', in which only site 1 is driven.  
\mbox{Destructive} two-photon interference leads to the effect seen in Fig.~\ref{fig:blockade}, which demonstrates that two photons never occur together in this site in the steady state, giving an excellent single photon source.

Using this example, we can also show how to calculate multi-time correlations with positive-P.  Any multi-time correlation function that is normally- and time-ordered can be calculated in the positive-P representation in a simple way, by averaging the corresponding product of phase space variables over the trajectories \cite{tcorr,QuantumNoise}. This follows by straightforward extension of the derivation found in Gardiner \cite{QuantumNoise} for the Glauber-P representation.  Such is not the case in the Truncated Wigner approach, which is based on symmetrically ordered operators, making computing useful time correlations challenging \cite{Berg09,Polkovnikov10}.
As an example, in Fig.~\ref{fig:g2tau}, we show the two time density correlations $g^{(2)}_{1,1}\!\left(\tau\right)$ of the driven site in the steady state:
\begin{equation}
g^{(2)}_{1,1}\!\left(\tau\right) = \frac{\langle\dagop{a}_1\!\left(t\right)\dagop{a}_1\!\left(t+\tau\right)\op{a}_1\!\left(t+\tau\right)\op{a}_1\!\left(t\right)\rangle}{\langle\dagop{a}_1\!\left(t\right)\op{a}_1\!\left(t\right)\rangle\langle\dagop{a}_1\!\left(t+\tau\right)\op{a}_1\!\left(t+\tau\right)\rangle}.
\end{equation}
In the positive-P representation this can be calculated as 
\begin{equation}\label{g1t}
g^{(2)}_{1,1}\!\left(\tau\right) = \frac{{\rm Re}\langle\alpha_1\!\left(t\right)\alpha_1\!\left(t+\tau\right)\wt{\alpha}^*_1\!\left(t+\tau\right)\wt{\alpha}^*_1\!\left(t\right)\rangle_s}{N_1\!\left(t\right)N_1\!\left(t+\tau\right)},
\end{equation}
where $N_1\!\left(t\right) = {\rm Re}\langle\alpha_1\!\left(t\right)\wt{\alpha}^*_1\!\left(t\right)\rangle_s$ as defined in \eqn{obs}, and the factors inside the numerator average in \eqn{g1t} are constructed using different time values coming from the same realization.  The form of $g^{(2)}_{1,1}\!\left(\tau\right)$ shows the characteristic oscillations with the delay $\tau$, as seen in previous literature on the unconventional photon blockade \cite{Liew10,Bamba11}.  

For this two site system, it is possible for us to compare to exact numerical solutions of the master equation.  It can be seen in both Figs. \ref{fig:blockade} and \ref{fig:g2tau}, that there is a strong agreement between the positive-P and more direct numerical integration of the master equation.

\subsection{Lieb Lattices}
\label{LIEB}

\begin{table*}[htb]
\begin{center}
\begin{tabular}{|cc|ccc|cccc|}
\hline
\multicolumn{2}{|c|}{configuration}	&	\multicolumn{3}{c|}{corner space renormalization\cite{Casteels16}}& \multicolumn{4}{c|}{positive-P}\\	
 \quad $N_{\rm cells}$ \quad & \quad $J/\gamma$ \quad & \quad $n_B/n_A$ \quad & \quad $g_B^{(2)}$ \quad & \quad $g^{(2)}_{B,\rm nn}$ \quad & \quad $n_B/n_A$ \quad & \quad $g_B^{(2)}$ \quad & \quad $g^{(2)}_{B,\rm nn}$ \quad & samples $\mc{S}$ \quad \\
\hline
 \quad $12$ \quad & \quad 2 \quad & \quad 0.0180(5) \quad & \quad 342(8) \quad & \quad 19.3(4) \quad & \quad 0.0176(4) \quad & \quad 342(16) \quad & \quad 19.0(5)	\quad & \quad $10^4$ \quad \\
 \quad $12$ \quad & \quad 1 \quad & \quad 0.0650(3) \quad & \quad 23.3(2) \quad & \quad 2.35(2) \quad & \quad 0.065(1) \quad & \quad 23(1) \quad & \quad 2.30(6) \quad & \quad 1000 \quad \\
 \quad $100$ \quad & \quad 1 \quad & \quad -- \quad & \quad -- \quad & \quad -- \quad & \quad 0.0648(2) \quad & \quad 23.3(2) \quad & \quad 2.36(4) \quad & \quad 1000 \quad \\
\hline
 \quad $4\times 4$ \quad & \quad 2 \quad & \quad 0.0161(1)	\quad & \quad 66.2(2) \quad & \quad 1.42(3) \quad & \quad 0.0161(3) \quad & \quad 65(2) \quad & \quad 1.2(2) \quad & \quad 1000 \quad \\
 \quad $4\times 4$ \quad & \quad 1 \quad & \quad 0.0631(1)	\quad & \quad 4.41(1) \quad & \quad 0.996(2) \quad & \quad 0.0628(3) \quad & \quad 4.42(3) \quad & \quad 0.99(2) \quad & \quad 1000 \quad \\
 \quad $10\times 10$ \quad & \quad 1 \quad & \quad -- \quad & \quad -- \quad & \quad -- \quad & \quad 0.0632(2) \quad & \quad 4.68(3) \quad & \quad 0.996(5) \quad & \quad 1000 \quad \\
 \quad $100\times 100$ \quad & \quad 1 \quad & \quad -- \quad & \quad -- \quad & \quad -- \quad & \quad 0.06309(8) \quad & \quad 4.685(2) \quad & \quad 0.995(2) \quad & \quad 100 \quad \\
\hline
\end{tabular}
\end{center}\vspace*{-0.5cm}
\caption{
Comparison between positive-P and corner space normalization calculations for the stationary state of 1d (top) and 2d (bottom) Lieb lattices.
Here, $J_{ij}=J$ for all connected sites, $U=0.3\gamma$,  $\Delta=0$, and $F_c=0.1\gamma$ in the driven sites (C sites only) with periodic boundary conditions.
$n_A$ and $n_B$ are the occupations of A and B sites, respectively, while $g_B^{(2)}$ is the on-site two body correlation on the B sites.
$g^{(2)}_{B,\rm nn}=\langle\dagop{a}_{B,j}\dagop{a}_{B,k}\op{a}_{B,j}\op{a}_{B,k}\rangle/n_B^2$ is the normalized density correlation between B sites in nearest neighbor unit cells $j$ and $k$.
\label{tab:dark}}
\end{table*}

\begin{figure}[htb]
\begin{center}
\includegraphics[width=\columnwidth]{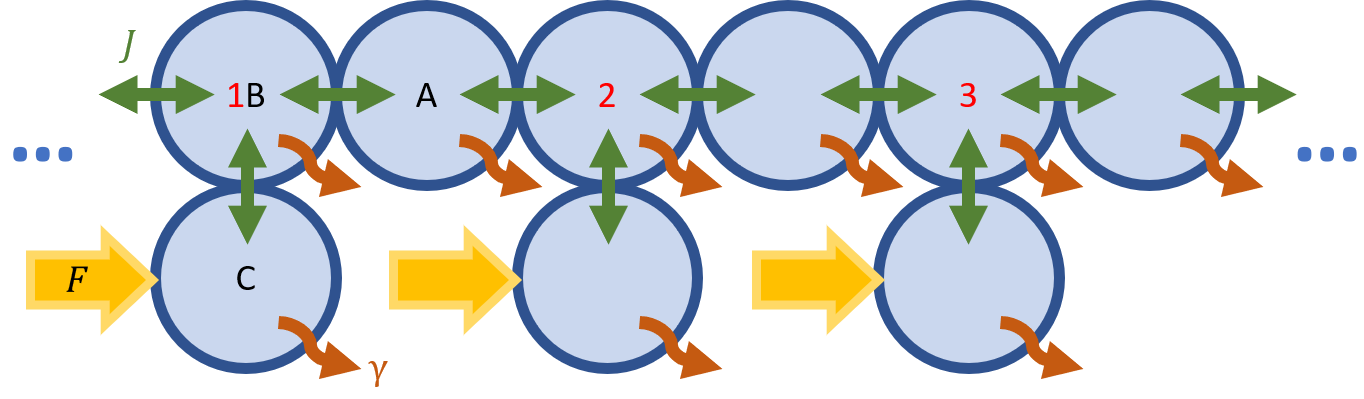}\\[3mm]
\includegraphics[width=0.8\columnwidth]{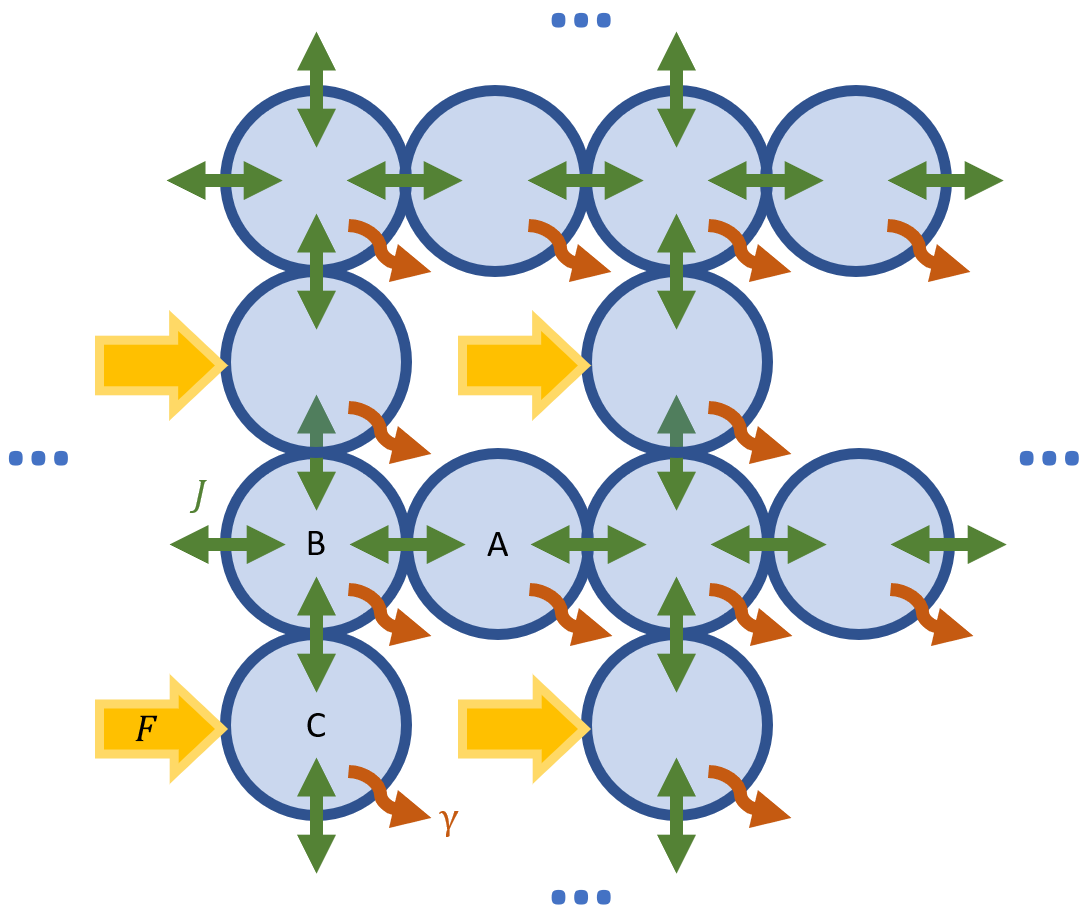}
\end{center}\vspace*{-0.5cm}
\caption{{\bf Lieb Lattices.}
A schematic of the 1d (top) and 2d (bottom) Lieb lattices, pumped on the C sites, resulting in dark B sites.
\label{fig:lieb}}
\end{figure}

Going beyond the two mode case to more complicated systems, we begin by considering the much studied case of a Lieb lattice \cite{Baboux16,Klembt17,Whittaker18,Goblot19}, which exhibits frustration and a flatband structure. The unit cells contain 3 sites (labeled A,B,C), and only some connections allow tunneling between cells, as per the schematic shown in Fig.~\ref{fig:lieb}.
A 1d Lieb lattice has been implemented e.g. by polaritons in an array of micropillars \cite{Baboux16,Goblot19}.
A Lieb lattice pumped locally only on the C sites has dark B sites that have far more striking departures from coherence than the single sites of Sec.~\ref{1MODE}  \cite{Casteels16}.  
In directly pumped sites, the field is usually close to being pinned by the coherent pump, whereas the dark sites are free to evolve to a much less classical stationary state.

The study of \cite{Casteels16} used the corner space renormalization method \cite{Finazzi15} to obtain accurate predictions for small lattice sizes, which provide a convenient benchmark for the precision and accuracy of the positive-P approach. 
Table~\ref{tab:dark} shows a comparison. 	
There is excellent agreement, and similar precision.
Due to the much more favorable scaling of the positive-P method, huge lattices are easily accessible.
Results for lattices with up to $100\times100$ unit cells are shown.  They indicate that for 1d, the 12 unit cell lattice saturates the infinite size limit. However, for the 2d system, the $4\times4$ lattice that was achievable by corner space renormalization does not yet reach the macroscopic limit in terms of density correlations $g^{(2)}$.  

\subsection{Uniform square lattices}

A uniform square lattice with tunneling 
between all nearest neighbor sites is also of much current interest. 
Here, we will use the notation of \cite{LeBoite13}, where $J_{ij} = J/z$ between nearest neighbor sites. The lattices have periodic boundary conditions and $M=m\times m$ sites in total. The coordination number is $z=4$ when $m>2$, 
and $z=2$ for the special case of $m=2$ in which left/right connections are to the same site.

The homogeneous case with uniform $F$, $U$, $J$, $\Delta$ has been studied using 
a self-consistent mean field (SCMF) approach 
pioneered by LeBoite\etal~ \cite{LeBoite13}. They found a flat-band, collective excitations, and a tunneling induced transition to bistability. Later work has also shown bimodality in the photon number distribution and a hysteretic cycle around a 1st order phase transition at higher tunneling \cite{Biondi17b}. 
The idea behind the SCMF is that the tunneling terms in the Hamiltonian can be expressed in the mean-field picture as 
\eq{scmf}{
-\frac{J}{z}\dagop{a}_i\op{a}_j \to -\frac{J}{z}\left(\langle\op{a}_i\rangle^*\op{a}_j + \langle\op{a}_j\rangle\dagop{a}_i\right),
}
which is equivalent to an effective coherent driving of
\eq{Fef}{
F_{\rm eff} = F - J\langle\op{a}\rangle.
} 
One then self-consistently solves for the exact quantum expressions from \cite{Drummond80b} for $\langle\op{a}\rangle$ in a single mode while using 
$F_{\rm eff}(\langle\op{a}\rangle)$ from \eqn{Fef} as the coherent driving. This can 
be done by iteration, starting with the bare $F$.
It is a similar approach to the self consistent mean-field widely used for conservative Bose-Hubbard models.  
Eq. \eqn{Fef} also lets one see that it may be useful to approximate coherent transport into the region of interest with an effective driving $F\approx-J\langle\op{a}\rangle$ in some systems.  
For cases with negligible quantum depletion, a symmetry broken ``Gross-Pitaevskii'' (GP) approach can also be used. This is equivalent to setting the quantum noises $\xi$ and $\wt{\xi}$ in \eqn{pp} to zero:
\eq{GP}{
\frac{\partial\alpha_j}{\partial t} = i\Delta_j\alpha_j -iU_j|\alpha_j|^2\alpha_j -iF_j -\frac{\gamma_j}{2}\alpha_j+\sum_{k} iJ_{kj}\alpha_{k}
}
so that the observable predictions (\ref{obs}) and (\ref{obs2}) reduce to 
$N = |\alpha|^2$ and  $g_2 = 1$.
Both approaches have evident gaps in the description: the SCMF assumes a uniform system 
(or potentially, a local density approximation), and does not take into account any spatial correlations. The GP approach can treat inhomogeneities properly, but does not take into account quantum depletion at all. 
How does the full quantum approach of the positive-P compare?

\begin{figure}[htb]
\begin{center}
\includegraphics[width=\columnwidth]{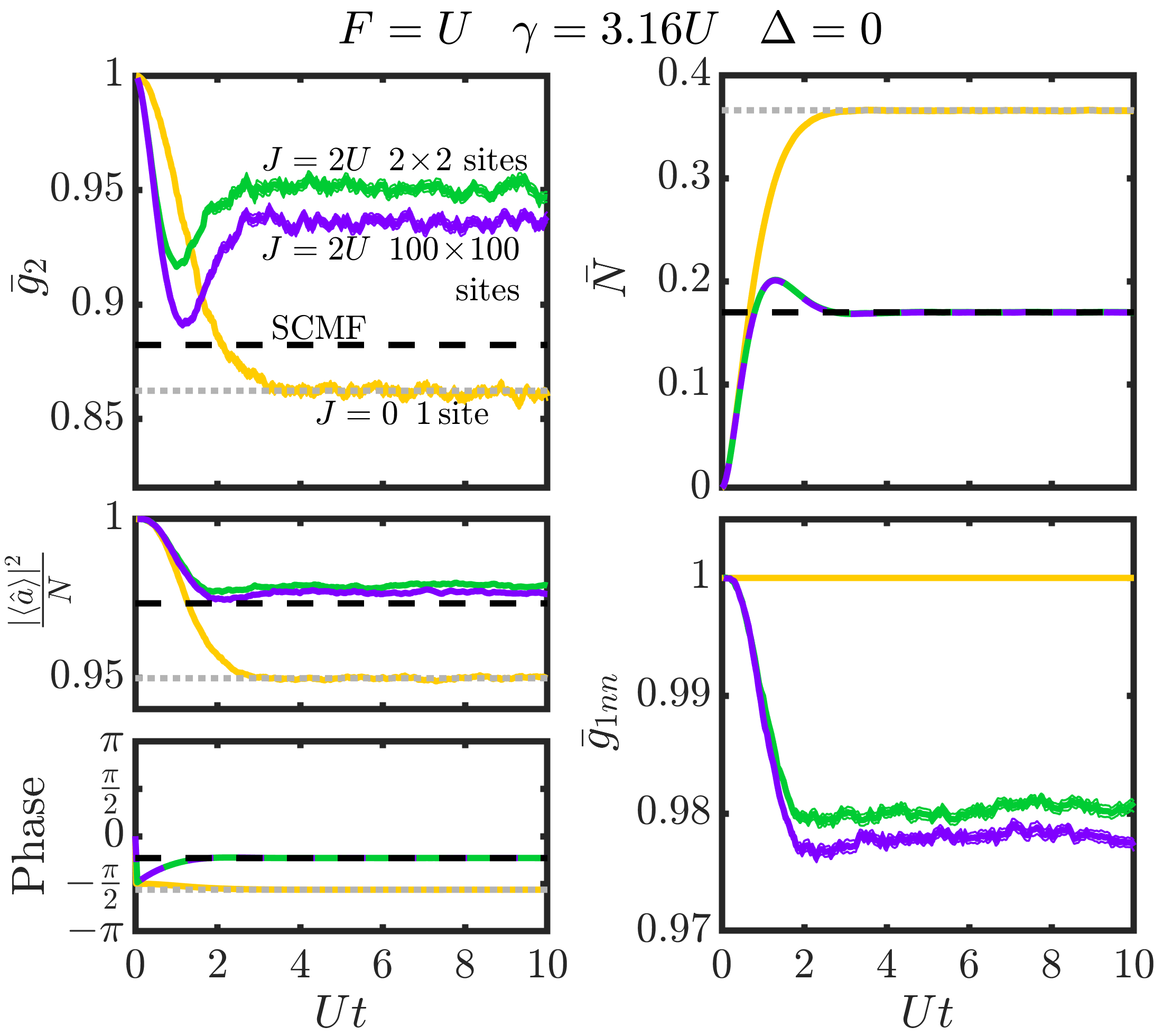}
\end{center}\vspace*{-0.5cm}
\caption{ {\bf Square Lattices.}
Simulations with different lattice sizes for $F=U$, $\gamma=3.16\,U$, $\Delta=0$ are shown: 
a $100\times100$ lattice with $J=2U$ (violet), 
a $2\times2$ lattice with $J=2U$ (green), and
a single site with $J=0$ (yellow).
Panels show:
$\wb{g}_2$ the density-density correlation $g_2$ averaged over all sites (top-left),
the average occupation per site $\wb{N}$ (top-right), the average amplitude $\langle\op{a}\rangle$ (bottom-left). In both the latter, the $J=2U$ cases overlap. Bottom-right: the nearest neighbor 1st order coherence (normalized) as given by \eqn{g1nn}. 
Also shown are 1-site exact values \cite{Drummond80b} (dotted) and the SCMF predictions \cite{LeBoite13} for many modes (dashed). 
\label{fig:J2}}
\end{figure}

\begin{table}[htb]
\begin{center}
\begin{tabular}{|l|cc|c|cc|}
\hline
		& \multicolumn{3}{c|}{estimates} 																			& \multicolumn{2}{c|}{full quantum}\\
\cline{2-6}
		& \multicolumn{2}{c|}{\parbox[t]{2.5cm}{single-mode \cite{Drummond80b}}} & \parbox[t]{1.5cm}{SCMF \cite{LeBoite13}} 	& \multicolumn{2}{c|}{positive-P}\\	
\cline{2-6}
$F =$ 	& $F_0$		& $F_0$		& $F_0-J\langle\op{a}\rangle$	&  $F_0$		& $F_0$		\\	
$\Delta =$	& $\Delta_0$	& $\Delta_0+J$	& $\Delta_0$			&  $\Delta_0$	& $\Delta_0$	\\
lattice 	& $1\times1$	& $1\times1$	& $1\times1$			& $2\times 2$	&  $100\times 100$\\ 
\hline
$\langle\dagop{a}\op{a}\rangle$
		& 0.3659		& 0.1750 		& 0.1701		& 0.17080(6)	& 0.17067(6)	\\	
$|\langle\op{a}\rangle|^2/N$			
		& 0.9498	& 0.9751		& 0.9734		& 0.9788(4)	& 0.9766(5)	\\	
arg$\langle\op{a}\rangle$
		& -1.7680		& -0.7333 		& -0.7219		& -0.7223(2)	& -0.7230(3)	\\	
$\wb{g}_2$		& 0.8624		& 1.4092		& 0.8824		& 0.9531(6)		& 0.9383(4)		\\	
$\wb{g}_{1\rm nn}$			
		& --			& --			& --					& 0.98943(3)	& 0.98823(4)	\\	
\hline
\end{tabular}
\end{center}\vspace*{-0.5cm}
\caption{Comparison of lattice values to estimates.
 $F=F_0=U$, $\gamma=3.16\,U$, $J=2\,U$, $\Delta=\Delta_0=0$.
Description in text. 
\label{tab:est}}
\end{table}

First, we consider the uniformly driven case with periodic boundary conditions. 
$F,U,\Delta$, and $J$ are identical at all sites. 
In Fig.~\ref{fig:J2}, positive-P simulations are shown for the crossover regime case of $F=U$, $\gamma=3.16U$, $\Delta=0$ studied in Fig.~\ref{fig:1m}(b), now with a nonzero tunneling $J=2U$ on small and large lattices. 
The 1 site case is also shown for reference in yellow. 
The move from single site to modest lattice to huge lattice is basically effortless in terms of calculation difficulty.
Numerical comparisons with standard estimates are shown in Table~\ref{tab:est}.
The quantity
\begin{equation}
\wb{g}_{1\rm nn} = \frac{\sum_{i,j}\langle\dagop{a}_{i,j}\op{a}_{i,j+1}\rangle}{M\wb{N}}
\label{g1nn}
\end{equation}
gives the average 1st order coherence between nearest neighbor sites, where $\wb{N}$ is the average occupation.

The first thing to note is that there is a significant influence of $J$: basically none of the observables agree between the 1-site model shown in yellow and the lattice calculations. 
Furthermore, the $2\times 2$ lattice is not sufficient to reach the asymptotic behavior, as seen in both 1st and 2nd order correlations. However, the mean amplitude and occupation can mislead one into thinking that the asymptotic limit has been reached. 
Since huge lattices of $100\times 100$ are easily accessible, a positive-P calculation can be used to determine the size required to reach the asymptotic regime.
In the case of the parameters of Fig.~\ref{fig:J2}, 
a $5\times 5$ lattice is needed for accurate convergence, as shown in Table.~\ref{tab:latt}.

\begin{table}[htb]
\begin{center}
\begin{tabular}{|c|cc|r|}
\hline
lattice \quad & \quad $\wb{g}_2$ \quad & \quad $\wb{g}_{1\rm nn}$ \quad & \quad samples $\mc{S}$ \quad \\
\hline\hline
 $2\times 2$ \quad & \quad 	0.9531(6) \quad  & \quad 0.98943(3) \quad & \quad 250\,000 \quad \\	
 $3\times 3$ \quad & \quad 	0.9327(3) \quad & \quad 0.98845(7) \quad & \quad 100\,000 \quad \\	
 $4\times 4$ \quad & \quad 	0.9372(4) \quad & \quad 0.98820(4) \quad & \quad 50\,000 \quad \\	
 $5\times 5$ \quad & \quad 	0.9386(5) \quad & \quad 0.98829(8) \quad & \quad 40\,000 \quad \\	
 $10\times 10$ \quad & \quad 	0.9389(6) \quad & \quad 0.98819(6) \quad & \quad 10\,000 \quad \\	
 $100\times 100$	\quad & \quad 	0.9383(4) \quad & \quad 0.98823(4) \quad & \quad 100 \quad \\
\hline
\end{tabular}
\end{center}\vspace*{-0.5cm}
\caption{
Lattice size scaling of correlations for $F=F_0=U$, $\gamma=3.16\,U$, $J=2\,U$, $\Delta=\Delta_0=0$. 
Notice that for a given precision, the number of samples decreases approximately proportionally to the lattice size.
\label{tab:latt}}
\end{table}

An important question is whether the mean-field approach is faithful to the asymptotic limit of many sites.
The SCMF is remarkably good for $\wb{N}$ and $\langle\op{a}\rangle$, but poor for density fluctuations $\wb{g}_2$. This potentially sheds some doubt on past results obtained this way \cite{LeBoite13,Biondi17b}, at least  in similar regimes.  
Going to even simpler approaches, a cut down version of the SCMF simply calculates the single-mode exact value with detuning modified as per $\Delta\to\Delta_{\rm eff}=\Delta+J$. This is also shown in Table.~\ref{tab:est}. We note that there is already a large improvement over the $J=0$ estimate, except for $\wb{g}_2$, which is sensitive to quantum correlations.
None of the estimates are able to give any information about $\wb{g}_{1\rm nn}$.

\subsection{Non-uniform pumping}

\begin{figure}[htb]
\begin{center}
\includegraphics[width=\columnwidth]{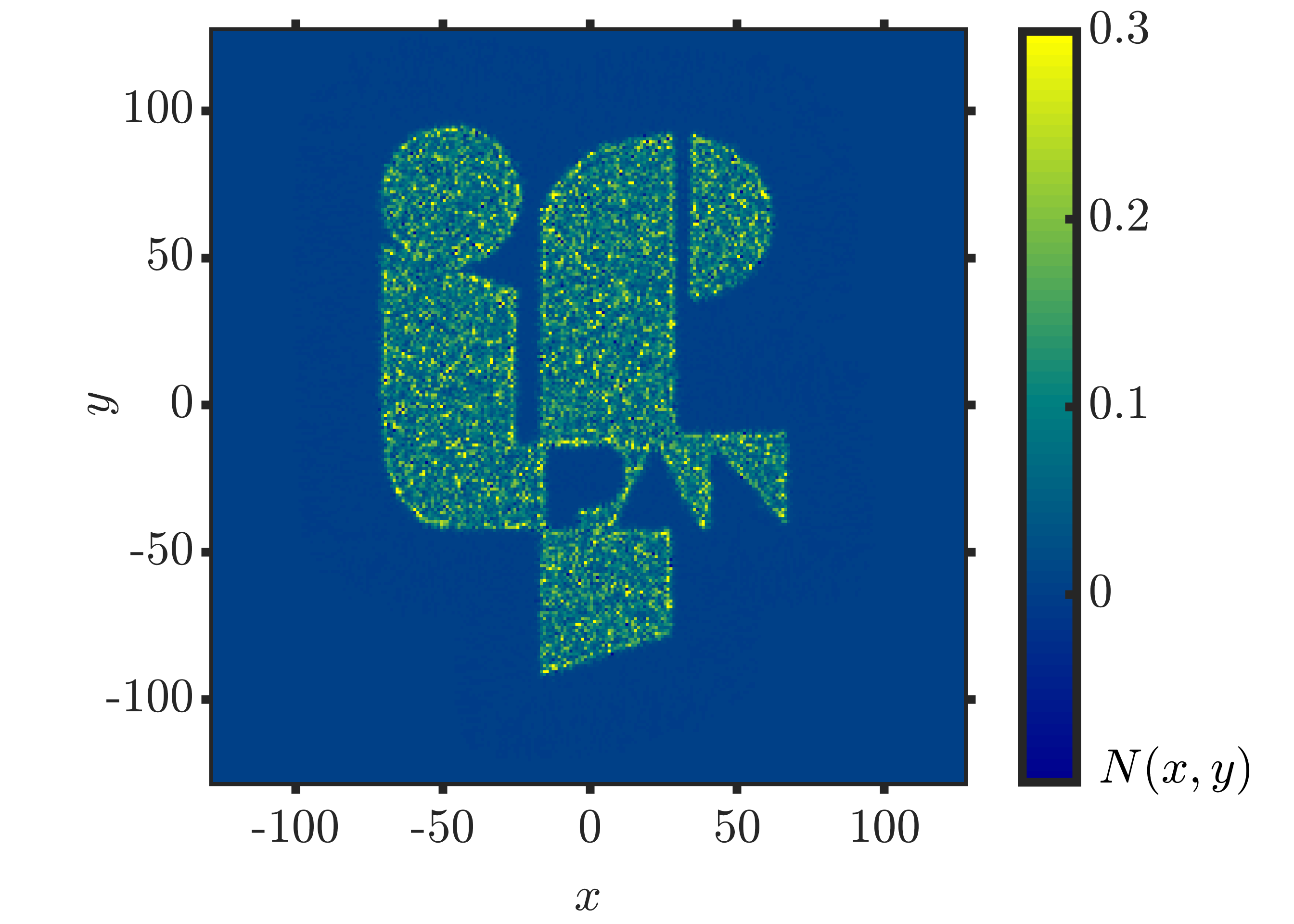}\\[2mm]
\includegraphics[width=\columnwidth]{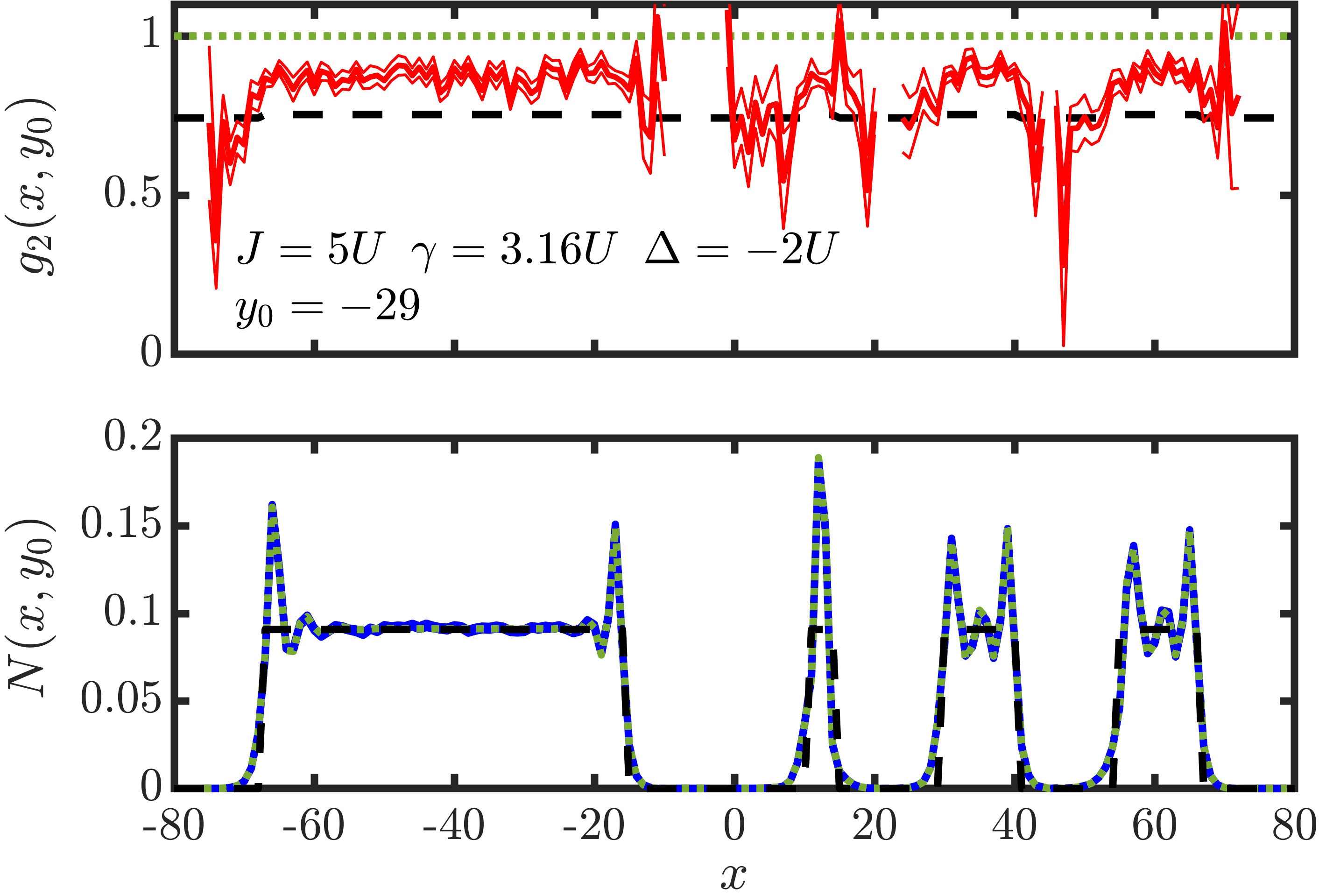}
\end{center}\vspace*{-0.5cm}
\caption{
{\bf Non-uniform driving.} A large $256\times 256$ site lattice, with parameters $J=5U$, $\gamma=3.16U$, $\Delta=-2U$. Local driving $F(x,y)=U$ or $F(x,y)=0$ according to the shape of the Institute of Physics logo.  
Top-panel: an instantaneous density in a snapshot of a single realization at the steady-state.
Bottom panels: steady-state observables along the line $y=-29$, calculated with $4000$ realizations. Solid lines: positive-P calculation of $g_2$ and $N$ with $1\sigma$ error bars; dashed lines: SCMF 
 predictions \cite{LeBoite13} based on the local value of $F$; dotted yellow lines: GP 
calculation using \eqn{GP}. 
\label{fig:ifpan}}
\end{figure}

A situation where particularly large lattices are necessary is when the system is nonuniform, or excitations involve many sites collectively \cite{Biondi17b,Fitzpatrick17,Naether15}. 
Systems of $10^4-10^6$ sites pose no problem for the positive-P approach, potentially allowing for complicated geometries, extensive transport, or simulations of emergent phenomena. 
Fig.~\ref{fig:ifpan} shows results for a truly large $256\times 256$ system with complicated geometry.

Spreading of $N(x,y)$ away from the pumped area is observed simultaneously with coherent spatial oscillations as a surface effect around the pumping zone. 
These behaviors are captured well by the Gross-Pitaevskii equation (\ref{GP}). 
The SCMF approach does not replicate the emergent local structure due to tunneling, though the bulk density is properly described.

On the other hand, the density fluctuations are not well described by either of the approximate methods -- only the positive-P gives an accurate description, even in the bulk. This last aspect is consistent with what we saw in Fig.~\ref{fig:J2} and Table.~\ref{tab:est}. At the points furthest from the driven region, $g_2$ seems to tend towards the SCMF estimate, though it becomes very noisy, as one would also expect in experiment, due to the very low density (e.g. observe the regions around $x=-70,5,25,50$, the furthest points from the driven region for which the occupation is still sufficient to have $g_2$ measurable beyond the noise).
Notably, the positive-P calculation allows one to predict the spatial variation of $g_2$ in the vicinity of the surface, which is not possible either accurately or even qualitatively by the approximate approaches.  

One realization of the simulation shown in Fig.~\ref{fig:ifpan} took 80s on a single PC processor (Intel Xeon E5645, 2.40GHz). Calculations on a $1000\times 1000$ lattice took 1h per realization under the same fairly basic conditions. The calculation time grows approximately linearly with $J$ for these parameters due to time step requirements.

\section{Nonzero temperature} 
\label{TEMP}

The master equation \eqn{drhodt} assumes dissipation into empty modes. 
A more general form is
\begin{eqnarray}\label{masterT}
\frac{\partial\op{\rho}}{\partial t} &=& -i\left[\op{H},\op{\rho}\right] +\sum_j\frac{\gamma_jN^B_j}{2}\left[2\dagop{a}_j\op{\rho}\op{a}_j -\op{a}_j\dagop{a}_j\op{\rho} - \op{\rho}\op{a}_j\dagop{a}_j\right] \nonu\\
&&+\sum_j \frac{\gamma_j(N^B_j+1)}{2}\left[2\op{a}_j\op{\rho}\dagop{a}_j -\dagop{a}_j\op{a}_j\op{\rho} - \op{\rho}\dagop{a}_j\op{a}_j\right],
\end{eqnarray}
which can be used to model systems coupled to baths with finite occupations $N^B_j$. 
The correction to the FPE of \eqn{ppFPE} is then
\begin{equation}
\frac{\partial P}{\partial t} =\eqn{ppFPE}_{RHS} + 
\sum_j\left[\frac{\partial^2}{\partial\alpha_j\partial\wt{\alpha}^*_j}+\frac{\partial^2}{\partial\wt{\alpha}^*_j\partial\alpha_j}\right]\frac{\gamma_jN^B_j}{2} P,
\label{newP}
\end{equation}
while the additions to the equations of motion \eqn{pp} are
\begin{eqnarray} 
\frac{\partial\alpha_j}{\partial t} &=&\eqn{ppa}_{RHS} + 
\sqrt{\gamma_jN^B_j}\,\eta_j(t)  \nonu \\
\frac{\partial\wt{\alpha}_j}{\partial t} &=&\eqn{ppb}_{RHS} + 
\sqrt{\gamma_jN^B_j}\,\eta_j(t)\label{eomT}
\end{eqnarray}
with \textsl{complex} white noises $\eta_j$ of mean zero that obey 
\eqa{eta}{
\langle\eta^*_j(t)\eta_{k}(t')\rangle_s &=& \delta(t-t')\delta_{jk} , \nonu \\
\langle\eta_j(t)\eta_{k}(t')\rangle_s &=& 0. 
}

\begin{figure}[htb]
\begin{center}
\includegraphics[width=\columnwidth]{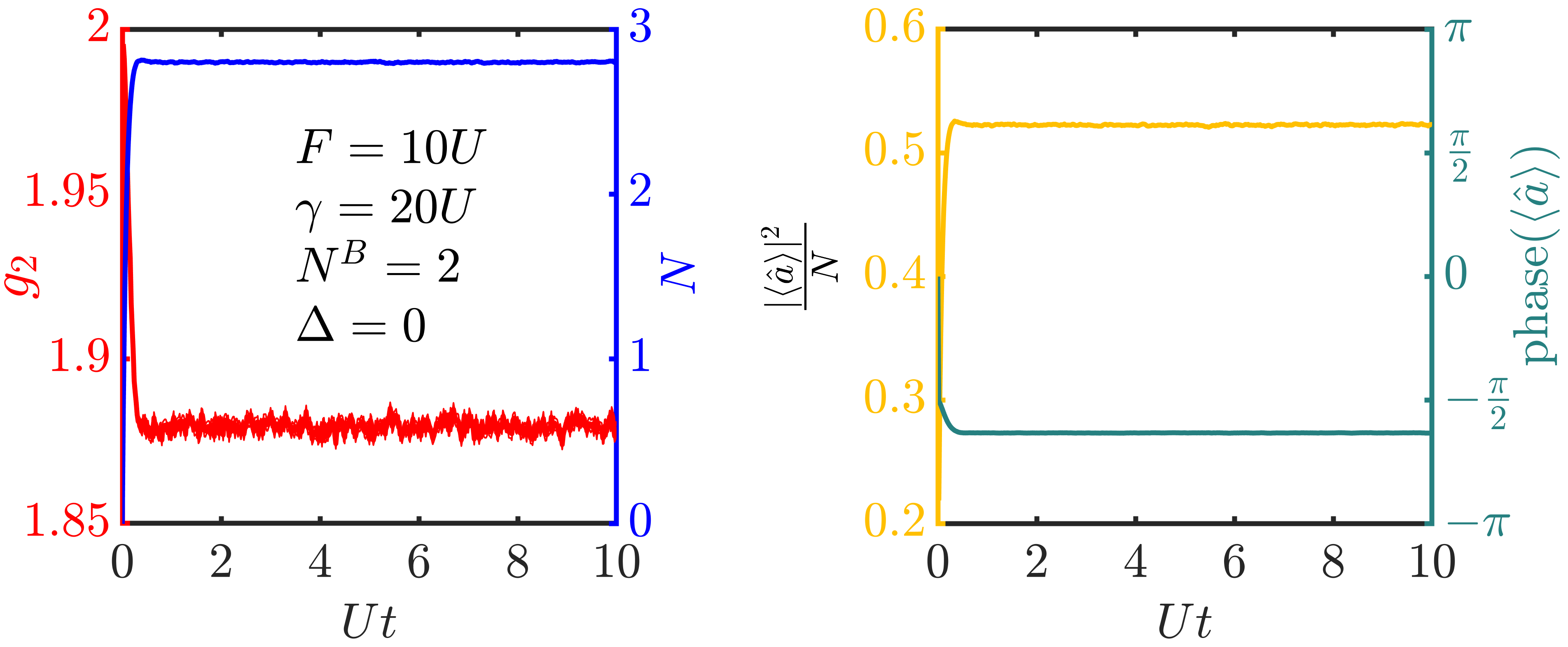}
\end{center}\vspace*{-0.5cm}
\caption{{\bf Finite temperature bath.} Single mode with dissipation into a bath with occupation $N^B = 2$.  Other parameters are $F = 10U$, $\gamma = 20U$, $\Delta = 0$.  
Steady state values in the $N^B=0$ system are: $N=0.9860$, $g_2=0.9886$, $|\langle\op{a}\rangle|^2/N=0.9953$, phase = $-0.5309\pi$.
\label{fig:N2}}
\end{figure}

In Fig. \ref{fig:N2}, we give an example of a single mode with coherent drive and decay into occupied modes.  Compared to the vacuum bath case, the steady state value of $g_2$ falls much closer to the value $g_2 = 2$ that would occur for thermal states; meanwhile, the coherence $\frac{|\left<\hat{a}\right>|^2}{N}$ falls much lower than in previous examples with dissipation to empty modes, showing that the positive-P method still works well for less coherent states. We thus expect the method to apply to condensates with low condensate fractions, materialization and other problems with no or weak coherence.

In the absence of coherent driving $F$, interactions $U$, and tunneling $J_{ij}$, Eq.(\ref{newP}) with the bath coupling leads to a stationary distribution of $P(\vec{v})=\otimes_jP_j$ with
\begin{equation}
P_j(\alpha_j,\wt{\alpha}_j) = \text{const.}\times \exp\left[-\frac{|\alpha_j|^2}{N^B_j}\right]\delta^{(2)}(\alpha_j-\wt{\alpha}_j).
\end{equation}
This is a thermal ensemble with occupations $n_j=|\alpha_j|^2$ and on average $N^B_j$ quanta at site $j$. 
The thermal occupation of each mode with energy $E_j=-\Delta_j$ can be considered as Bose distributed $N^B_j=\{\exp[(E_j-\mu)/k_BT]-1\}^{-1}$ in which $T$ and $\mu$ are resultant effective parameters of the reservoir, and in equilibrium -- also of the system.

When both the tunneling and the temperature are appreciable, so that density fluctuations become important, a proper treatment of the coupling of the system to the reservoir should involve the extended single particle states, instead of the local (site) basis. 
 For a Markovian reservoir, where particle and energy exchange takes place through interaction between system and reservoir quanta, a model that has often been used \cite{Gardiner03,Stoof99,Duine01,Wouters09,Rooney12,nocut} replaces local $N_j^B$ in \eqn{masterT} with an effective Bose-Einstein distributed occupation $N^B\to \{\exp[(\op{H}-\mu\op{N})/k_BT]-1\}^{-1}$.
A thermal bath of this kind in the relatively high temperature limit $\exp[(E-\mu)/k_BT]\to 1+(E-\mu)/k_BT$ has been implemented using the positive-P method for the closely related continuum ultracold atom systems \cite{Swislocki16}. They differ from our Hamiltonian
 by having kinetic energy rather than site-to-site tunneling, and lacking coherent driving. 
Since these terms  describe one-particle processes, their contribution to the stochastic differential equations is obtained by simply replacing  $\op{a}_j\to\alpha_j$ and $\dagop{a}_j\to\wt{\alpha}^*_j$ in the Heisenberg equations of motion. 
Hence for our driven-dissipative model  the corresponding equations become:
\eqa{sgpe-like}{
\frac{\partial\alpha_j}{\partial t} &=& \left(-i-\frac{\Gamma}{2}\right)\left\{(U_j\alpha_j\wt{\alpha}^*_j -\Delta_j)\alpha_j +F - \sum_{k} J_{kj}\alpha_{k} \right\}\nonu\\
&&+ \sqrt{-iU_j(1-i\Gamma)}\,\alpha_j\,\xi_j(t) + \sqrt{\Gamma T}\,\eta_j(t), \nonu \\
\frac{\partial\wt{\alpha}_j}{\partial t} &=& \left(-i-\frac{\Gamma}{2}\right)\left\{(U_j\wt{\alpha}_j\alpha_j^*-\Delta_j)\wt{\alpha}_j +F - \sum_{k} J_{kj}\wt{\alpha}_{k}\right\}\nonu\\
&&+ \sqrt{-iU_j(1-i\Gamma)}\,\wt{\alpha}_j\,\wt{\xi}_j(t) +\sqrt{\Gamma T}\,\eta_j(t),
}
with a reservoir at temperature $T$ ($k_B=1$), and coupling constant $\Gamma$. The conditions of applicability of \eqn{eomT} and \eqn{sgpe-like} are different, though there is an overlap regime. In that regime one can identify the correspondences $\Gamma=\gamma_j N^B_j/(k_B T)$ and $N^B_j=k_BT/(E_j-\mu)$, with $\mu$ incorporated into the $\Delta_j$.
Space-dependent temperature profiles can be 
included through a site-dependence of $T$.  

A rigorous derivation and consideration of applicability criteria for the equations \eqn{sgpe-like} goes beyond the scope of the article, but we include them for completeness of the picture regarding thermal effects.
We also mention that thermal baths that take into account the quantum particle-like nature at high energies have been implemented in \cite{Wouters09,nocut}.

\section{Conclusions}
\label{CONC}

We have described the essential elements for applying the positive-P method to 
the driven dissipative Bose-Hubbard model, and benchmarked its accuracy -- confirming lack of systematics down to the 4th significant digit in our test cases. 
The method appears to be a versatile and robust way to describe the full quantum mechanics of even very large systems, allowing for the study of the stationary state -- and its time correlations -- provided that the dissipation is sufficiently strong. 

In particular, one can reach strong antibunching (Figs.~\ref{fig:1m}(c)), high driving and occupation (Figs.~\ref{fig:1m}(a)), and crossover regimes, and even a strong photon blockade with perfect antibunching and destructive interference (Fig.~\ref{fig:blockade}). Large non-uniform systems with $256\times256$ sites (Fig.~\ref{fig:ifpan}) and more are easily accessible, opening the way to the study of much uncharted territory: e.g. dissipative phase transitions in a flat-band with large scale fluctuations \cite{Biondi15,Fitzpatrick17,Biondi17b}, or the point at which bimodality predicted by semiclassical approaches \cite{LeBoite13,LeBoite14} morphs to a 1st order phase transition \cite{Casteels17,Fink17}. 
It can also be readily used to determine the minimal sizes of systems required to reach the asymptotic regime.  We have studied Lieb lattices (Table~\ref{tab:dark}) and simple orthogonal 2d lattices (Table.~\ref{tab:latt}, Fig.~\ref{fig:J2}) in this regard, showing that $4\times4$ systems e.g. tend not to be in the asymptotic limit.

The method exhibits clear superiority over various mean field approaches and the truncated Wigner approximation in the difficult regime when occupation is small, and provides the ability to simultaneously and accurately study correlations, interference, tunneling and nonlocal effects. Due to the numerical instabilities, the positive-P approach cannot describe strongly driven weakly damped conditions. However, this more semiclassical regime is extremely well approximated by the related truncated Wigner method with (from the technical point of view) very similar stochastic equations to be solved.  
Thus, between truncated Wigner, which gives very accurate results for large occupations, and positive-P, we have a viable method for all conditions where either drive or dissipation are significant effects.  Notably, the positive-P approach gives full quantum results in the medium to large dissipation regime, whereas most other full quantum approaches such as DMRG, tensor networks etc. work more easily under the opposite, low dissipation, conditions.

In the usable regime, numerical effort scales merely linearly with the number of sites, and quadratically with the precision.
Space and time-dependence of all parameters in the model is easily incorporated with no extra numerical effort. Nonlocal interactions can also be efficiently treated \cite{Wuster17}.
Thus we suggest positive-P as the method of choice to access large systems in the very regions that are currently experimentally relevant, especially in driven-dissipative but correlated photonic platforms.  

Due to the additional stability provided by dissipation, the positive-P is applicable to a much wider range of problems in open dissipative systems than in closed systems. 
Its great success in describing the archetypal driven dissipative Bose-Hubbard model shown here implies that positive-P may be an ideal method for simulating various kinds of open quantum systems that either consist of or can be mapped onto bosons.  
Such promising extensions, such as incoherent driving or systems with coupled spins and bosons, will be the subject of future work.

\acknowledgments
We are grateful to 
Filippo Gaggioli, Sebastian Schmidt,
Peter Drummond,
Ashton Bradley,
Elena Ostrovskaya,
and 
Jacqueline Bloch
for inspiring discussions. 
M.H.S. gratefully acknowledges financial support from QuantERA InterPol and EPSRC (Grant No. EP/R04399X/1 and No. EP/K003623/2).
M.M. acknowledges support from the National Science Center, Poland via grant No. 2017/25/Z/ST3/03032, under the QuantERA program, P.D. from grant No. 2018/31/B/ST2/01871.
This work was granted access to the HPC resources of CINES under the allocations 2019-A0060507629 and 2020-A0080507629 supplied by GENCI (Grand Equipement National de Calcul Intensif).

\bibliography{artnew}

\appendix

\section{Derivation of Positive-P equations}
\label{DevPP}

Here we cover the mathematical details behind the derivation of the positive-P method.
A central element upon which the following derivations are based are the differential identities
\begin{eqnarray}
\op{a}_j\op{\Lambda}_j &=& \alpha_j\op{\Lambda}_j, \nonu \\
\dagop{a}_j\op{\Lambda}_j &=& \left[\wt{\alpha}^*_j+\frac{\partial}{\partial\alpha_j}\right]\op{\Lambda}_j, \nonu \\
\op{\Lambda}_j\op{a}_j &=& \left[\alpha_j+\frac{\partial}{\partial\wt{\alpha}^*_j}\right]\op{\Lambda}_j, \nonu \\
\op{\Lambda}_j\dagop{a}_j &=& \wt{\alpha}^*_j\op{\Lambda}_j.
\end{eqnarray}
These allow one to convert expectation values of observables and the evolution equation \eqn{drhodt} to functions of only the variables $\vec{v}$ and the distribution $P$, leaving $\op{\Lambda}_j$ as the only remaining operators.
For example, the expectation value of the site occupation is
\eqs{Nj}{
N_j&=&\langle\dagop{a}_j\op{a}_j\rangle = \Tr{ \dagop{a}_j\op{a}_j\op{\rho} }  \label{Nj1}\\
&=& \int d^{4M}\vec{v}\ P(\vec{v})\,\Tr{ \dagop{a}_j\op{a}_j\op{\Lambda}(\vec{v}) }\label{Nj2}\\
&=& \int d^{4M}\vec{v}\ P(\vec{v})\,\alpha_j\left[\wt{\alpha}^*_j+\frac{\partial}{\partial\alpha_j}\right]\Tr{ \op{\Lambda} }\quad\\
&=& \int d^{4M}\vec{v}\ P(\vec{v})\,\alpha_j\wt{\alpha}^*_j\label{Tr1}\\
&=& \lim_{\mc{S}\to\infty}\langle \alpha_j\wt{\alpha}^*_j \rangle_s.\label{Njend}
}
The line \eqn{Tr1} follows from ${\rm Tr}[\op{\Lambda}_j]=1$, since any derivative of $1$ is zero.
Notably, evaluating \eqn{Njend} with reasonable precision (say, 3-4 significant digits) can be far more efficient than evaluating the trace with the full density matrix in \eqn{Nj1} when the system is large.
This is where the power of the method comes from.

In similar vein, to obtain an evolution equation for the samples $\vec{v}$, the master equation \eqn{drhodt} is first converted to an integral equation of the form
\eqa{inteq}{
\lefteqn{\int d^{4M}\vec{v}\ \op{\Lambda}\ \frac{\partial P}{\partial t}=} &&   \\
&&\qquad\int d^{4M}\vec{v}\ P\,\Bigg\{ \sum_v A_v(\vec{v})\,\frac{\partial}{\partial v} 
+ \sum_{vv'} \frac{D_{vv'}(\vec{v})}{2}\frac{\partial^2}{\partial v\partial v'} \Bigg\} \op{\Lambda},\nonu
}
with $v,v'$ denoting variables in $\vec{v}$, and the coefficients $A$ and $D$, making a form akin to \eqn{Nj2}. 
This step and subsequent ones have been explained in detail for the present system in \cite{DeuarPhD}. 

The RHS of \eqn{inteq} can be integrated by parts to give derivatives of $P$ instead of $\op{\Lambda}$, plus boundary terms at $|\alpha|,|\wt{\alpha}|\to\infty$, which are discarded. i.e. 
\eqa{intbt}{
\lefteqn{\int d^{4M}\vec{v}\ \op{\Lambda}\,\frac{\partial P}{\partial t} =}&&\\
&&\qquad \int d^{4M}\vec{v}\ \op{\Lambda} \Bigg\{ - \frac{\partial}{\partial v} \sum_v A_v(\vec{v})
+ \sum_{vv'} \frac{\partial^2}{\partial v'\partial v}\frac{D_{vv'}(\vec{v})}{2} \Bigg\} P.\nonu
}
The discarding of boundary terms relies on an assumption of self-consistency: That as long as the distribution (and therefore the sample trajectories) are well behaved, that is $P \to 0$ sufficiently fast as $\alpha,\wt{\alpha}\to \infty$, the boundary terms are zero, and vice versa. 
Indeed, for poorly damped interacting systems there can be a time $t_{\rm sim}$ around which divergent trajectories or huge excursions appear, indicating that the move from \eqn{inteq} to \eqn{intbt} is failing from this time onward \cite{Gilchrist97}. In such a case, results at subsequent times $t>t_{\rm sim}$ should be discarded.  This is now a well studied and controlled element of the theory, and it is known that the once feared ``boundary term bias'' \cite{Smith89,Kinsler91} becomes obscured by noise before it can affect results \cite{Gilchrist97}.  In Bose-Hubbard like models, a noise amplification that masks meaningful results appears just prior to $t_{\rm sim}$ \cite{Gilchrist97,Deuar06a,Deuar06b,Swislocki16,Wuster17}.

An equation like \eqn{intbt} of the form $\int d^{4M}\vec{v}\ \op{\Lambda}\,f(\vec{v}) = 0$ has potentially many solutions, but one of them certainly is $f(\vec{v})=0$. 
Therefore, from \eqn{intbt} one obtains a Fokker-Planck equation for $P$. In our model, the equation is 
\eqa{ppFPE}{
\frac{\partial P}{\partial t} = 
-\sum_j&&\frac{\partial}{\partial\alpha_j}\left\{ -i\left[F_j+U_j\alpha_j^2\wt{\alpha}_j^*-\Delta_j\alpha_j\right] -\frac{\gamma_j}{2}\alpha_j\right\} P \notag  \\
-\sum_j&&\frac{\partial}{\partial\wt{\alpha}_j^*}\left\{ i\left[F_j^*+U_j\wt{\alpha}_j^{*2}\alpha_j-\Delta_j\wt{\alpha}^*_j\right] -\frac{\gamma_j}{2}\wt{\alpha}_j^*\right\} P \notag \\ 
+\sum_j&&\frac{\partial^2}{\partial\alpha_j^2}\frac{(-iU_j)}{2}\,\alpha_j^2 P +\sum_j\frac{\partial^2}{\partial\wt{\alpha}^{*2}_j}\frac{iU_j}{2}\,\wt{\alpha}^{*2}_j P \notag \\
-\!\!\!\!\!\!\!\! \sum_{{\rm connections}\,j,k} &&\left[ \frac{\partial}{\partial\alpha_j}iJ_{kj}\alpha_{k} + \frac{\partial}{\partial\alpha_{k}}iJ_{jk}\alpha_j \right. \notag \\
&&\left. - \frac{\partial}{\partial\wt{\alpha}^*_j}iJ_{kj}^*\wt{\alpha}^*_{k} - \frac{\partial}{\partial\wt{\alpha}^*_{k}}iJ_{jk}^*\wt{\alpha}^*_j\right]\,P.
}
This gives the $\left(\alpha_j, \wt{\alpha}^*_j\right)$ components of the drift $A$ and diffusion $D$ as
\eqa{ADcomp}{
A_j &=& \begin{pmatrix}  \left[i\Delta_j-\frac{\gamma_j}{2}-iU_j\alpha_j\wt{\alpha}_j^*\right]\alpha_j -iF_j +\sum_{k} iJ_{kj}\alpha_{k} \nonu \\ 
\left[-i\Delta_j-\frac{\gamma_j}{2}+iU_j\wt{\alpha}_j^{*}\alpha_j\right]\wt{\alpha}_j^* +iF_j^*  - \sum_{k} iJ^*_{kj}\wt{\alpha}^*_{k} \end{pmatrix} ,  \\
D_{jj'} &=&  \begin{pmatrix} (-iU_j)\alpha_j^2 & 0 \\ 0 & iU_j\wt{\alpha}^{*2}_j \end{pmatrix} .}
Fokker-Planck equations with non-negative diffusion can be converted to stochastic differential equations by standard methods \cite{QuantumNoiseNew,StochMech}. The form of the kernel $\op{\Lambda}$ which is analytic in the complex variables $\alpha_j$ and 
$\wt{\alpha}_j^*$ allows one to always ensure that the diffusion is non-negative through a standard transformation (see \cite{Drummond80,DeuarPhD} for detail of the procedure).   For a Fokker-Planck equation with drift vector  $A$ and diffusion matrix $D$, the corresponding set of stochastic differential equations is
\eq{stocheq}{\frac{\partial \vec{v}}{\partial t} = A(\vec{v}) + B(\vec{v})\vec{\xi}(t) , 
}
where $\vec{\xi}(t)$ are uncorrelated real Gaussian white noises with zero mean and \mbox{$\langle\xi_v(t)\xi_{v'}(t')\rangle_s = \delta(t-t')\delta_{vv'}$}; the matrix B is such that $D = BB^T$ and is generally non-unique for a given $D$.  For the diffusion matrix $D$  in \eqn{ADcomp}, we chose a matrix $B$ whose $\left(\alpha_j, \wt{\alpha}^*_j\right)$ components are:
\eq{Bcomp}{
B_{jj'} = \begin{pmatrix} \sqrt{-iU_j}\alpha_j & 0 \\ 0 & \sqrt{iU_j}\wt{\alpha}^*_j \end{pmatrix} .
}
This leads to the form of the stochastic equations given in \eqn{pp}.

\section{Useful simulation times}
\label{SIMT}

It is known \cite{Deuar06a} that for a closed undamped system, the noise catastrophe does rear its head, around the time 
\eq{tsim}{
t_{\rm sim} \approx \left\{\begin{array}{l@{\quad{\rm if}\quad}l}\frac{2.5}{\max_j[ U_j N_j^{2/3}]}& {\rm max}_j\,N_j\gg1 \\ \frac{C}{{\rm max}_j\,U_j} & {\rm max}_j\,N_j\ll1 \end{array}\right.
}
where $C\sim 10$ is a numerical constant. The estimate \eqn{tsim} is borne out qualitatively in our simulations.
The basic trade-off has been that while results for short evolution times are always accessible, a nonlinear amplification of the trajectory spread would eventually appear at long times and obscure predictions below a rising noise floor. 
Dissipation has been shown to stabilize the positive-P equations above a threshold strength \cite{Gilchrist97}.
A later study of simulation times $t_{\rm sim}$ introduced a characteristic logarithmic variance 
$\mc{V} = {\rm var}\left[\log|\alpha|+\log|\wt{\alpha}|\right]/2$ that can not exceed $\mc{O}(10)$ for a useful signal-to-noise ratio \cite{Deuar06a}.
A simplified version of the medium-time estimates made there for a single mode gives
\begin{eqnarray}
\lefteqn{\mc{V}\approx \frac{Ut}{2}+}&&\\
&& + U^2N^2\left[\frac{1}{q-\gamma}\left(\frac{1-e^{-\gamma t}}{\gamma}+\frac{e^{-qt}-1}{q}\right) -\frac{\left(1-e^{-\gamma t}\right)^2}{2\gamma^2}\right],\nonu
\end{eqnarray}
where $q=2(\gamma-U)$. This is suggestive that, at least for large $N$ (when the 2nd line is dominant), the variance growth is arrested if $q>0$, that is $\gamma >U$ (because then all exponentials are decaying with $t$).
For small $N$, on the other hand, we have $\mc{V}\approx Ut/2$. Notice now that the time to reach the stationary state must be at least several times $1/\gamma$ (say, 6 times). Hence to reach this without first breaking the $\mc{V}\sim10$ limit, at the least one needs  $3U/\gamma \ll 10$, i.e. $\gamma\gg U/3$.
In both cases, the regime $\gamma\gtrsim U$ looks promising for simulations that make it into the stationary state.
However, this has not been actually tested in numerical calculations prior to the current work.

For the coherently driven dissipative model we consider in this work, the steady state does not depend on the initial conditions chosen. 
While one could in principle choose an initial state with significantly larger occupation than the steady state, and hence large $\alpha$ and $\widetilde{\alpha}$, the self-amplification of the noise terms could cause the simulation to fail earlier. Such dynamical effects have been extensively studied in \cite{Deuar06a,Deuar06b}, with the conclusion that stability is essentially determined by the maximum occupation during evolution. Based on \eqn{tsim}, this only arises when $N_j\gg1$. For the above reasons, we choose and recommend a vacuum initial state as a simple universal option for reaching the stationary state that will not cause such unnecessary instabilities in cases that would otherwise be stable.

\section{Stability diagram for nonzero $\Delta$}
\label{ADELTA}

\begin{figure}[htb]
\begin{center}
\includegraphics[width=\columnwidth]{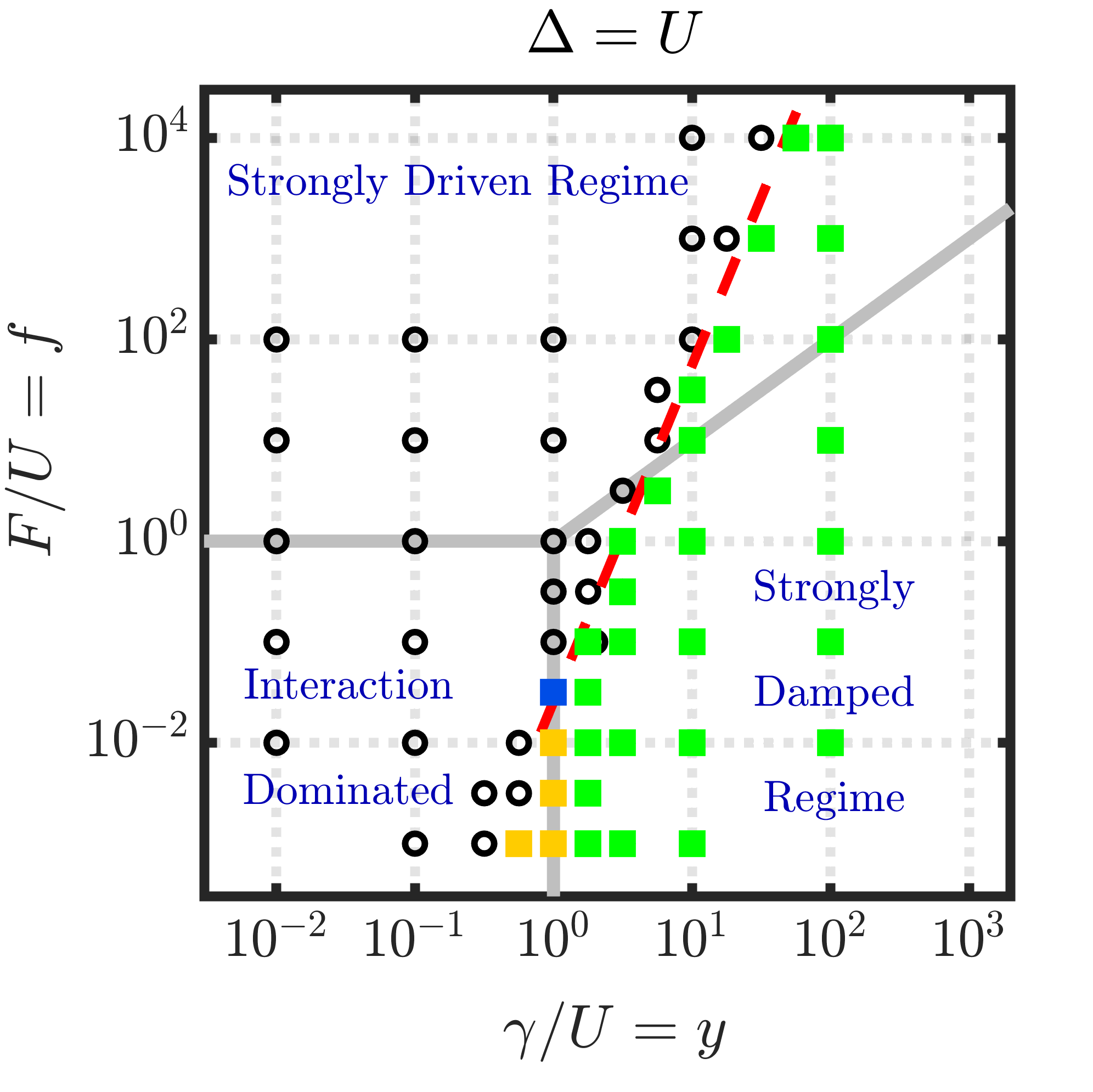}
\end{center}\vspace*{-0.5cm}
\caption{{\bf Regimes of usefulness of positive-P for \mbox{$\Delta=U$}}. All notation the same as in Fig.~\ref{fig:phasediag}, apart from the change in $\Delta$. 
\label{fig:phasediagdw1}}
\end{figure}

As seen in Sec.~\ref{DETUNING} and elsewhere \cite{Biondi17b}, the natural energy scale for $\Delta$ is $U$. 
The usefulness diagram of positive-P simulations is shown in Fig.~\ref{fig:phasediagdw1} for the relevant case of $\Delta=U$. This indicates that detuning does not introduce large modifications to the picture already seen in Fig.~\ref{fig:phasediag}, or the expressions \eqn{usability}, at least on a log-log scale.  

\section{Truncated Wigner equations and errors}
\label{WIG}

The evolution equations in the truncated Wigner representation, corresponding to \eqn{pp}, are
\eqa{Wig-eq}{
\frac{\partial\alpha_j}{\partial t} &=& i\Delta_j\alpha_j -iU_j(|\alpha_j|^2-1)\alpha_j -iF_j -\frac{\gamma_j}{2}\alpha_j \nonu\\
&& + \sqrt{\frac{\gamma_j}{2}}\,\eta_j(t)+\sum_{k} iJ_{kj}\alpha_{k},
}
with complex white noise $\eta$ as per \eqn{eta}. In principle, one should start with half a particle's worth of complex noise in each mode as per $\alpha_j(0) = \chi_j/\sqrt{2}$, where 
\begin{equation}
\langle \chi^*_j\chi_{k}\rangle_s=\delta_{jk};\qquad
\langle \chi_j\chi_{k}\rangle_s=0. 
\end{equation}
However, the stationary state does not depend on this, because the dynamical noise generates the appropriate variance (provided the truncation error is small). 
Observable predictions use the ensemble average of the Weyl symbols:
\begin{eqnarray}
N &=& \langle|\alpha|^2\rangle_s -\frac{1}{2},\\
g_2 &=& \frac{\langle |\alpha|^4 - 2|\alpha|^2 + \frac{1}{2}\rangle_s}{N^2}.
\end{eqnarray}

As mentioned in section \ref{WIGCOMP}, we assess the accuracy of our TW simulations for the single site problem 
using the estimates of the four observables $N$, $g_2$, $|\langle \op{a}\rangle|^2/N$ and phase $\arg\langle\op{a}\rangle$.
For given values of  the parameter ratios $\gamma/U$ and $F/U$, we carry out TW calculations using $s=96$ subensembles, each containing $10416$ trajectories. Next, we extract the best estimates $O^{(j)}\pm \delta_{\rm stat}O^{(j)}$, with $j=1,..,4$, for the 
four observables, to be compared with the exact predictions $O_{\rm ex}^{(j)}$ by Drummond and Walls \cite{Drummond80b}.
For each observable we compute the systematic and statistical \emph{relative} errors as
\begin{equation}\label{rel}
\Delta_{\rm sys}^{(j)}=\left |\frac{O^{(j)}-O_{\rm ex}^{(j)}}{O_{\rm ex}^{(j)}}\right |;\;\;\;\; \Delta_{\rm stat}^{(j)}=
\frac{\delta_{\rm stat}O^{(j)}}{|O^{(j)}|}.
\end{equation}

\begin{figure} [htb]
\begin{center}
\includegraphics[width=\columnwidth]{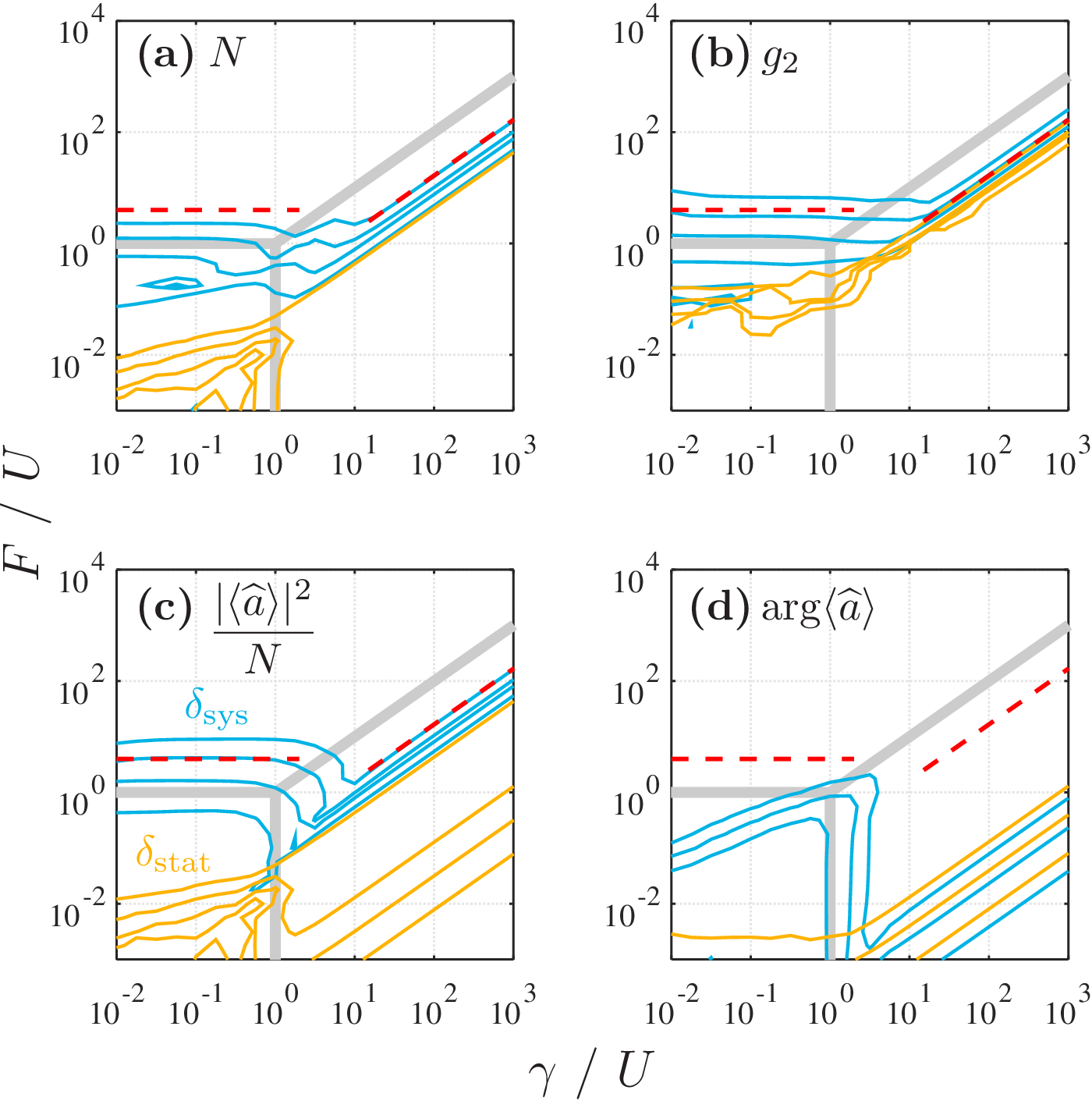}
\end{center}
\caption{{\bf Error budgets in the truncated Wigner method for different observables}.  Blue lines show contours of the systematic error $\Delta_{\rm sys}$ at values of $0.01$, $0.03$, $0.1$, and $0.3$ (top to bottom), for the main observables. The same contours for the statistical error $\Delta_{\rm stat}$ with $\approx10^6$ realizations are shown in yellow. Red dashed lines are the overall limits \eqn{urulewig}. All data is for zero detuning ($\Delta=0$). 
\label{fig:wigerr}}
\end{figure}

We then repeat the same procedure by varying the ratios $\gamma/U$ and $F/U$ over several orders of magnitude, using a grid of size $21\times29$. 
Fig.~\ref{fig:wigerr} shows the results obtained for the systematic and statistical relative errors (\ref{rel}) for the four observables.  
We see that the systematic error is the stronger restriction in practically all cases. In particular, there are very large regions in which this error is seen without being masked by attendant statistical error. 
We also notice that at low $\gamma$ the highest systematic error comes from the coherent amplitude characterized by $|\langle\op{a}\rangle|^2/N$, whereas at high $\gamma$ the limiting systematic error is from $g_2$, though the corresponding errors in $N$ and $|\langle\op{a}\rangle|$ are also comparable.  
We finally take the largest relative error
\eq{DeltaTW}{
\Delta_{TW}={\rm max}_j [\Delta_{\rm sys}^{(j)}, \Delta_{\rm stat}^{(j)}]
}
as the overall assessment of the errors expected from the truncated Wigner.  This quantity is displayed in Fig.~\ref{fig:TW}.  From the above discussion, its behavior closely follows the systematic relative errors in either the coherent amplitude 
$|\langle\op{a}\rangle|^2/N$ or in the $g_2$.

\end{document}